\documentclass[conference]{IEEEtran}
\usepackage{ifthen}
\newif\ifshort
\shortfalse %

\ifshort
\newcommand{\isShort}{true}
\else
\newcommand{\isShort}{false}
\fi
\newcommand{\shortVer}[1]{\ifthenelse{\equal{\isShort}{true}}{{#1}}{}}
\newcommand{\longVer}[1]{\ifthenelse{\equal{\isShort}{false}}{{#1}}{}}

\usepackage{protocol}

\usepackage[marginal,hang,stable]{footmisc}
\renewcommand{\footnoterule}{%
  \kern -3pt
  \hrule width 1in 
  \kern 2pt
}
\usepackage{graphicx}
\graphicspath{ {figures/} }

\usepackage{amssymb}
\usepackage{amsmath}
\usepackage{colortbl}
\usepackage{color}
\usepackage{amsthm}

\usepackage{flushend}

\usepackage[small,bf]{caption}

\renewcommand{\figurename}{Figure}

\usepackage{times}
\usepackage{indentfirst}
\usepackage{url}
\usepackage[labelformat=simple]{subcaption}

\usepackage{xspace}
\usepackage{multirow}
\definecolor{darkgreen}{RGB}{47,109,79}
\definecolor{darkblue}{RGB}{57,79,99}
\usepackage[bookmarks=false, colorlinks=true, plainpages = false,  linkcolor = darkgreen,   citecolor = darkgreen, urlcolor = darkblue, filecolor = blue]{hyperref}
\usepackage{threeparttable}
\usepackage{tikz}
\usepackage{bbding}
\usepackage[all]{xy}
\usepackage{array}
\usepackage{multirow}
\usepackage{paralist}
\usepackage{xspace}
\usepackage{url}
\usepackage{graphicx}
\usepackage{latexsym}
\usepackage{amssymb}
\usepackage{amsfonts}
\usepackage{psfrag}
\usepackage{wrapfig}
\usepackage{comment}
\usepackage{color, colortbl}

\DeclareRobustCommand{\ttfamily}{\fontfamily{lmtt}\selectfont}

\newtheoremstyle{mystyle}%
  {}%
  {}%
  {}%
  {}%
  {\bfseries\itshape}%
  {.}%
  { }%
  {}%
\theoremstyle{mystyle}

\renewcommand{\scriptsize}{\fontsize{7.5}{9.5}\selectfont}

\makeatletter
\def\url@leostyle{%
  \@ifundefined{selectfont}{\def\UrlFont{\small\ttfamily}}%
  {\def\UrlFont{\scriptsize\ttfamily}}%
}
\makeatother	
\newcommand{\descr}[1]{\medskip \noindent \textbf{#1}}
\newcommand{\descrit}[1]{\smallskip \noindent \emph{#1}}

\newtheorem{definition}{Definition}

\usepackage{hhline}

\newcommand{\Zq}{\mathbb{Z}_q}
\newcommand{\Ui}{\mathcal{U}_i}
\newcommand{\Uon}{\mathcal{U}^{on}}

\newcommand{\BigO}[1]{\ensuremath{O(#1)}}

\newcommand{\user}{{\tt user}\xspace}
\newcommand{\users}{{\tt users}\xspace}
\newcommand{\User}{{\tt User}\xspace}
\newcommand{\Users}{{\tt Users}\xspace}
\newcommand{\tally}{{\tt tally}\xspace}

\newcommand{\Tally}{{\tt Tally}\xspace}

\usepackage[square,sort,comma,numbers]{natbib}

\begin{document}

\sloppy

\title{Efficient Private Statistics with Succinct Sketches}

\author{\IEEEauthorblockN{Luca Melis, George Danezis, and Emiliano De Cristofaro}
\IEEEauthorblockA{Department of Computer Science, University College London\\ \{luca.melis.14, g.danezis, e.decristofaro\}@ucl.ac.uk}}

\thispagestyle{plain}

%\IEEEoverridecommandlockouts
%\makeatletter\def\@IEEEpubidpullup{9\baselineskip}\makeatother
%\IEEEpubid{\parbox{\columnwidth}{Permission to freely reproduce all or part
%    of this paper for noncommercial purposes is granted provided that
%    copies bear this notice and the full citation on the first
%    page. Reproduction for commercial purposes is strictly prohibited
%    without the prior written consent of the Internet Society, the
%    first-named author (for reproduction of an entire paper only), and
%    the author's employer if the paper was prepared within the scope
%    of employment.  \\
%    NDSS '16, 21-24 February 2016, San Diego, CA, USA\\
%    Copyright 2016 Internet Society, ISBN 1-891562-41-X\\
%    http://dx.doi.org/10.14722/ndss.2016.23175
%}
%\hspace{\columnsep}\makebox[\columnwidth]{}}

\maketitle

\begin{abstract}
Large-scale collection of contextual information is often essential in order to gather statistics, train machine learning models, and extract  knowledge from data. The ability to do so in a {\em privacy-preserving} way -- i.e., without collecting fine-grained user data -- enables a number of additional computational scenarios that would be hard, or outright impossible, to realize without strong privacy guarantees. 
In this paper, we present the design and implementation of practical techniques for privately gathering statistics from large data streams. We build on efficient cryptographic protocols for private aggregation and on data structures for succinct data representation, namely, Count-Min Sketch and Count Sketch. These allow us to reduce the communication and computation complexity incurred by each data source (e.g., end-users) from linear to logarithmic in the size of their input, while introducing a parametrized upper-bounded error that does not compromise the quality of the statistics.
We then show how to use our techniques, efficiently, to instantiate real-world privacy-friendly systems, supporting recommendations for media streaming services, prediction of user locations, and computation of median statistics for Tor hidden services. 
\end{abstract}

\section{Introduction}
The increasing amount of contextual information collected by multitudes of always-on, always-connected devices makes it increasingly possible to extract value and knowledge from statistical data. For instance, Google analyzes GPS locations reported by mobile devices to calculate the speed along a road and generate live traffic maps (Google Traffic), and search data to estimate and predict flu activity (Google Flu Trends). %
Alas, the large-scale collection of user data raises serious privacy, confidentiality, and liability concerns. 
This motivates the need for efficient and scalable techniques allowing providers to {\em privately} gather statistics, and to use such statistics to train models and facilitate predictions. Our work is actually inspired by a few real-world problems:
\begin{enumerate}
\setlength\itemsep{0.5em}
\item[\bf P1]  Online streaming services routinely collect statistics about videos watched by their users, and provide them with personalized suggestions, typically, using recommender systems. 
In particular, we will focus on recommendations for BBC's iPlayer~\cite{iplayer}, an online platform offering free streaming of TV programs.
\item[\bf P2] Urban planning committees, as well as mass transport operators, are keen on gathering statistics about movements and commuting paths, aiming to improve transportation services and predict future trends, e.g., to respond to anomalies and disruptions on short notice~\cite{scellato2011nextplace,pejovic2013anticipatory}.
\item[\bf P3] The Tor network~\cite{Tor} needs to collect traffic statistics such as the number of, and traffic generated by, hidden services, in order to fine tune design decisions and convince their funders of the value of the network~\cite{elahi2014privex}. 
\end{enumerate}

In general, we are interested in scenarios where providers need to train models based on aggregate statistics gathered from many data sources, and our goal is to do so without disclosing fine-grained information about single sources. 
In theory, we could turn to existing cryptographic protocols for privacy-friendly aggregation: using homomorphic encryption or secret sharing untrusted aggregators can collect encrypted readings but only decrypt the sum~\cite{chan2012privacy,Mobiquitous05,Kursawe:2011,
erkin2012private,shi2011privacy,bilogrevic2014s}. However, these tools require each data source to perform a number of cryptographic operations, and transmit a number of ciphertexts, {\em linear} in the size of their input, which makes them impractical when sources contribute large streams. 
For instance, in scenario P1, 
we need to collect distributions of ``co-views'' (i.e., pairs of videos watched by the same user) in order to perform recommendations based on K-Nearest Neighbor (KNN) algorithms~\cite{cover1967nearest}: even when only hundreds of programs are available, each user would have to encrypt and transmit a matrix of hundreds of thousands of values. 

Also, differential privacy could be used to let aggregators add noise to datasets so that other parties may perform statistical queries while the probability of identifying single records is minimized~\cite{cormode2012differentially}. However, differential privacy alone would not protect the privacy of single data sources w.r.t. the aggregators themselves. Although recent work such as RAPPOR~\cite{erlingsson2014rappor} supports, via input perturbation, differentially-private statistics collection with an untrusted aggregator, it actually requires millions of users in order to obtain reasonably accurate answers.

Our insight is to combine privacy-preserving aggregation with data structures supporting succinct data representation, namely,  {\em Count-Min Sketch}~\cite{cormode2005improved} and  {\em Count Sketch}~\cite{charikar2002finding} (introduced in Section~\ref{sec:cms}). Private aggregation is performed over the sketches, rather than the raw inputs. Despite an upper-bounded error in the aggregate is introduced, this allows us to reduce communication and computational complexity (for the cryptographic operations) from {\em linear} to {\em logarithmic} in the size of the inputs. We then use the resulting private statistics tools to instantiate protocols and build systems addressing applications P1--P3 discussed above, where the error does not affect the overall quality of the computation.

More precisely, in Section~\ref{sec:service}, we present a privacy-preserving recommender system allowing online streaming services like BBC's iPlayer to support recommendations without tracking their users. Users' browsers encrypt and transmit a  succinct representation of the co-view matrix (i.e., pairs of videos they have watched) so that the BBC can only decrypt the aggregate matrix (i.e., how many users have watched a given pair). This is broadcast back to the users and used to derive recommendations.
Next, in Section~\ref{sec:smartphone}, we introduce an Android application  enabling  users to report to a service provider their locations over time, in a privacy-preserving way, i.e., so that only aggregate statistics are disclosed.
We then show that these can be used to train a model geared to predict future movements.
Finally, in Section~\ref{sec:Tor}, we build a system for privately computing statistics of Tor hidden services, aiming to address the conflict between the importance to  collect (and publish) such statistics and the risk of  harming the privacy of individual Tor users. 
This  addresses an open problem raised by the Tor Project~\cite{goulethidden}. We show how to estimate median statistics by collecting an encrypted frequency distribution of the statistics across all Hidden Services Directories (HSDir).

We also discuss real-world deployment and present full-blown implementations of our techniques, in JavaScript, Android, and/or Python. Our  design makes it extremely easy for anyone to integrate our techniques -- as simple as installing a package from a public repository. User-side deployment is transparent too, as client-side code can run in the browser (in JavaScript), thus requiring no additional software to be installed or technical understanding of the cryptographic layer.

Our techniques are not limited to one particular model: on the contrary, we can support different trust, robustness, and deployment models. Although our three applications all gather statistics via private sketch aggregation, they do differ in a few key characteristics.  The privacy-friendly recommendation and location prediction systems (cf.~Section~\ref{sec:service}--\ref{sec:smartphone}) build atop a privacy-preserving aggregation scheme where private keys sum up to zero~\cite{Kursawe:2011,shi2011privacy}, and use the aggregator itself as a bulletin board to distribute users' public keys. 
We implement them in JavaScript to support seamless web application deployment and portability to multiple browsers as well as Android.
On the other hand, our first-of-its-kind protocol for median statistics of Tor hidden services (cf.~Section~\ref{sec:Tor}) uses additively homomorphic threshold decryption, relying on a set of non-colluding authorities. It is developed in Python so that it can be  integrated on Tor Hidden Service Directories. We also show how to integrate differential privacy guarantees by adding noise to leaked {\em intermediate} values during the median estimation process which does not involve non-linear operations.

\descr{Paper organization.} The rest of the paper is organized as follows.
Next section introduces relevant background information, then, 
Section~\ref{sec:service} and Section~\ref{sec:smartphone} present, respectively, a privacy-preserving recommender system for online broadcasters and an Android-based private location prediction service. Section~\ref{sec:Tor} introduces a system for privately computing the median statistics of Tor hidden services.
After reviewing related work in Section~\ref{sec:work}, the paper concludes with Section~\ref{sec:conclusion}.

\section{Preliminaries}
\label{sec:preliminaries}

\subsection{Cryptographic Background}
\label{sec:crypto}

\descr{Computational Diffie Hellman Assumption.}
Let $\mathbb{G}$ be a cyclic group of order $q$ ($\vert q \vert = \tau$, for  security parameter $\tau$), with generator $g$.
We say that the Computational Diffie Hellman (CDH) problem is hard if, for any probabilistic polynomial-time algorithm $\mathcal{A}$ and random $x , y$ drawn from $\mathbb{Z}_q$: \vspace{-0.08cm}
\begin{equation*}
\Pr \left[ \mathcal{A} ( \mathbb{G} , q , g, g^x , g^y ) = g^{xy} \right] \vspace{-0.08cm}
\end{equation*}
is negligible in the security parameter $\tau$.

\descr{Decisional Diffie Hellman Assumption.}
Let $\mathbb{G}$ be a cyclic group of order $q$ ($\vert q \vert = \tau$), with generator $g$.
We say that the Decisional Diffie Hellman (DDH) problem is hard if, for any probabilistic polynomial-time algorithm $\mathcal{A'}$ and random $x , y, z$ drawn from $\mathbb{Z}_q$: \vspace{-0.08cm}
\begin{equation*}
\resizebox{\columnwidth}{!}{
$\Big\vert \Pr \left[ \mathcal{A'} ( \mathbb{G} , q , g, g^x , g^y , g^z ) = 1 \right] - 
\Pr \left[ \mathcal{A'}( \mathbb{G},q,g,g^x,g^y, g^{x y} ) = 1 \right] \Big\vert$ \vspace{-0.08cm}
}
\end{equation*}
is negligible in the security parameter $\tau$.

\descr{Pairwise Independent Hash Functions.}
Let $H$ be a family of {\em random-looking} 
hash functions mapping values from a domain [D] to a range [R].
$H$ is  {\em pairwise independent} iff $\forall x \neq y \in [D]$ and 
$\forall a_1, a_2 \in [R]$: %
$ 
\Pr_{h \in H} \left[ h(x) = a_1 \wedge h(y) = a_2 \right] = \frac{1}{R^2}.  %
$

\subsection{Count-Min Sketch and Count Sketch}
\label{sec:cms}

\descr{Count-Min Sketch~\cite{cormode2005improved}} is a data structure that can be used to provide a succinct sublinear-space representation of multi-sets. An interesting property is that they enable aggregation of the multi-sets represented by two or more sketches using a linear operation on the sketches themselves. Prior uses of Count-Min Sketch include summarizing large amounts of frequency data for sensing, networking, natural language processing, and database applications~\cite{site}.

\vspace*{0.1cm}
\begin{definition}[\itshape  Count-Min Sketch]\label{def:count} 
A Count-Min Sketch with parameters $(\epsilon, \delta)$ is a two-dimensional array (table) $X$, with width $w$ and depth $d$. 
Given parameters $(\epsilon, \delta)$, set $d=\lceil \ln{T/\delta} \rceil$ and $w=\lceil e / \epsilon \rceil$, where $T$ is the number of items to be counted. %
Each entry of the table is initialized to zero. Then, $d$ hash functions
$ h_j:\lbrace{0,1\rbrace}^* \rightarrow \lbrace{0,1\rbrace}^w$, %
are chosen uniformly at random from a pairwise-independent family $\mathcal{H}$.\vspace*{-0.15cm}
\end{definition}

\descrit{Update Procedure.} To update item $i$ by a quantity $c_i$, $c_i$ is added to one element in each row, where the element in row $j$ is determined by the hash function $h_j$. The update is
denoted as $(i , c_i)$.
More precisely, to update the count for item $i$ to $c_i\in\mathbb{N}$, for each row $j$ of $X$, set: \vspace{-0.1cm} 
\[
 X[j, h_j(i)] \leftarrow X[j, h_j(i)] + c_i
\]

\descrit{Estimation Procedure.}
To estimate the count $\hat{c_i}$ for item $i$, we take the minimum of the estimates of $c_i$ from every row of $X$:\vspace{-0.1cm} 
\[
 \hat{c_i} \gets \min_{j} X[j, h_j(i)] \vspace{-0.15cm}
\]

\descrit{Error Upper Bound.} Given estimate $\hat{c_i}$, it holds: \begin{compactenum}
\item $c_i \leq \hat{c_i} $
\item $\hat{c_i} \leq c_i + \epsilon \sum_{j=1}^{T} |c_j|$  with probability $1-\delta$.
\end{compactenum}
(where $c_i$ is the true counter).

\descr{Count Sketch~\cite{charikar2002finding}} is a data structure which provides an estimate for an item's frequency in a stream. Count Sketch has the same update procedure as Count-Min Sketch, but differs in the estimation.
Specifically, given the table $X$ built on the stream, the row estimate of $c_i$ (which is the counter of item $i$) for row $j$ is computed based on two buckets: $X[i,h_j(i)]$ and $X[i,h'_j(i)]$, where $h'_j(i)$ is defined as: \vspace{-0.15cm}
\begin{equation*}
h'_{j}(i) := \begin{cases}
				h_j(i) - 1 & \text{if } h_j(i) \bmod 2 = 0 \\
				h_j(i) + 1 & \text{if } h_j(i) \bmod 2 = 1 \\
			  \end{cases} %
\end{equation*}
The estimate of $c_i$ for row j is then $$\left( X[j,h_j(i)] - X[j,h'_j(i)] \right)$$
To estimate the count $\hat{c_i}$ for item $i$, we take the median of the estimates of $c_i$ from every row of $X$:\vspace{-0.1cm} 
\[
 \hat{c_i} \gets \underset{j}{\mathit{median}} \left( X[j,h_j(i)] - X[j,h'_j(i)] \right) %
\]
Both Count-Min and Count Sketch are linear: the element-wise sum of the sketches representing two multi-sets yields the sketch of their union.

\subsection{Differential Privacy}
\label{sec:diffPrivacy}

Differentially private mechanisms allow a party publishing a dataset to make sure that only a bounded amount of information is leaked. Output perturbation mechanisms modify a statistic on a dataset $D$, prior to its release, using a randomized algorithm $A$, so that the output of $A$ does not reveal too much information about any particular row in $D$. %

\begin{definition}[{\em $\epsilon$-Differential privacy}~\cite{dwork2006differential}]
A randomized algorithm $A$ satisfies $\epsilon$-differential privacy, if for any two neighbor datasets $D_1$ and $D_2$ that differ only in one row, and for any possible output $R$ of $A$, it holds: 
\[ 
\Pr \left[ A(D_1) = R \right] \leq e^{\epsilon} \cdot \Pr \left[ A(D_2) = R \right]  \vspace{-0.2cm}
\]
\end{definition}
Note that $\epsilon$ here is used differently than in the Count-Min Sketch's definition. Although this somewhat overloads the notation for $\epsilon$, it is actually clear from the context if it relates to the data structure or to the differential privacy setting.

\descr{Laplace Mechanism.} In Section~\ref{sec:Tor}, we use the differentially private Laplace mechanism \cite{dwork2006calibrating}, which perturbs the output of a function $F$. Given $F$, the Laplace mechanism transforms $F$ into a differentially private algorithm, by adding independent and identically distributed (i.i.d.) noise (denoted as $\eta$) into each output value of $F$. The noise $\eta$ is sampled from a Laplace distribution $Lap(\lambda)$ with the following probability density function: $Pr[\eta = x] = \frac{1}{2\lambda} e^{−|x|\lambda}$.
Dwork~\cite{dwork2006differential} proves that the Laplace mechanism ensures $\epsilon$-differential privacy if $\lambda \geq \frac{S(F)}{\epsilon}$, with  $S(F)$ denoting the sensitivity of $F$,
defined as:\vspace{-0.1cm}
$$S(F) = \max\limits_{D_1,D_2}{|| F(D_1) - F(D_2) ||}_1 \vspace{-0.1cm}$$
where ${|| \cdot ||}_1$ denotes the L1 norm, and $D_1$ and $D_2$ are any two neighbor datasets.
Intuitively, $S(F)$ measures the maximum possible change in $F$'s output when we modify one arbitrary row in $F$'s input.

\subsection{ItemKNN-based Recommender Systems}
\label{sec:recSystems}
Recommender systems are used to predict the utility of a certain item for a particular user, based on their previous ratings as well as those of other ``similar'' users~\cite{resnick1997recommender}.
Consider a set of $N$ users and a list of $M$ items:
for each user, a rating can be associated to each item, based, e.g., on the user's explicit opinion about the item (e.g., 1 to 5 stars) or by implicitly deriving it from purchase records or browser history.

Machine learning can be used to predict the expected rating of an unrated item for a given user. The \emph{K-Nearest Neighbor (KNN)} classification algorithm %
finds the top-$K$ nearest neighbors for a given item, so that ratings associated with these are combined to predict unknown ratings.
In this paper, we use a variant called \emph{ItemKNN}~\cite{sarwar2001item}. The algorithm is trained using an item-to-item similarity matrix (correlation matrix), where each element expresses the similarity between a pair of items, and the Cosine Similarity is computed between vectors of items (e.g., user ratings for each item).

If ratings are binary values (e.g., viewed/not viewed), as in one of our applications (see Section~\ref{sec:service}), the Cosine Similarity between items $a$ and $b$ is:
\begin{equation}
\label{eq:sim}
\left\{ Sim \right\}_{ab} = \dfrac{C_{ab}}{\sqrt{C_a \cdot C_b}}
\end{equation}
where $C_{ab}$, $C_{a}$, and $C_{b}$ denote, respectively, the number of people who rated both $a$ and $b$,  $a$, and $b$.
Given the similarity matrix, we can identify the nearest neighbors for each item as the items with the highest correlation values. The final model then consists of the identity of the nearest neighbors and their correlation values (or \emph{weights}) which are used in the prediction process, i.e., the items that should be recommended. 

Note that, with ItemKNN, given the item-to-item matrix, each user could independently compare their ratings with the nearest neighbors of each item in the model. Upon finding a match, the weight is added to the prediction score for that item. The items are then ranked by their prediction scores and the top $K$ are taken as recommendations.

\subsection{Exponential Weighted Moving Average (EWMA)}
\label{sec:timePred}

Exponential Weighted Moving Average (EWMA) models~\cite{soldo2010predictive} can predict future values based on past values weighted with exponentially decreasing weights toward older values.
Given a signal over time $r(t)$, we indicate with $\tilde{r}(t + 1)$ the predicted value of $r(t + 1)$ given the past observations, $r(t')$, at time $t' \leq t$. 
Predicted signal $\tilde{r}(t + 1)$ is estimated as: \vspace{-0.15cm}
$$\tilde{r}(t + 1) = \sum\limits_{t'=1}^{t} \alpha (1-\alpha)^{t - t'} r(t') \vspace{-0.15cm}$$
where $\alpha \in (0, 1) $ is the smoothing coefficient, and $t' = 1, \dots, t$
indicates the training window, i.e., $1$ corresponds to the oldest observation while $t$ is the most recent one.

\vspace{3mm}

In the rest of this work, we present efficient techniques to estimate, in a private and distributed way, the training datasets required for ItemKNN-based Recommender System, Exponential Weighted Moving Average (EWMA) modeling, as well as median and other frequency statistics. The mechanisms combine traditional linear aggregation with sketches, for efficiency, and, when needed, differential privacy to limit information leakage.

\section{Private Recommender Systems For\\ Streaming Services}
\label{sec:service}
Media streaming services are becoming increasingly popular as numerous dedicated providers (e.g., Netflix, Amazon, Hulu) as well as ``traditional'' broadcasting services (e.g., BBC, CNN, Al-Jazeera) offer digital access to TV shows, movies, documentaries, and news.
One of the providers' goals is often continuous user engagement, thus, new content should periodically be suggested to users based on their interests. 
These recommendations are usually provided by means of {\em recommender systems}~\cite{herlocker2004evaluating, adomavicius2005toward} like ItemKNN (cf.~Section~\ref{sec:recSystems}), which typically require the full availability of users' ratings, whereas, we focus on a model where a provider like the BBC provides recommendations to its users, e.g., on iPlayer, {\em without tracking} their preferences and viewings. Note that iPlayer does not actually require users to register or have an account, which further motivates the need to protect users' privacy.

\subsection{Overview}
\label{subsec:secret}
We present a novel privacy-friendly recommender system where the ItemKNN algorithm is trained using only aggregate statistics. Aiming to build a global matrix of co-views (i.e., pairs of programs watched by the same user) in a privacy-preserving way, we rely on (i) private data aggregation based on secret sharing (inspired by the ``low overhead protocol'' in~\cite{Kursawe:2011}), and (ii) the Count-Min Sketch data structure to reduce the computation/communication overhead, trading off an upper-bounded error with increased efficiency.

Recommendations are derived, based on ItemKNN, as follows: users' interests are modeled as a (symmetric) item-to-item matrix $I = \lbrace{ 0, 1 \rbrace}^{M \times M}$, where $I_{ab}$ is set to $1$ if the user has watched both  programs $a$ and $b$ and to $0$ otherwise.
$I_{aa}$ is set to $1$ if the user has watched the program $a$.
The Cosine Similarity $\left\{ Sim \right\}_{ab}$ between programs $a$ and $b$ can be computed from item-to-item matrices using Equation~\ref{eq:sim}.
The Cosine Similarity is then used by each user to derive personalized recommendations as described in Section~\ref{sec:recSystems}. %

\descr{System Model.} Our system involves a \tally (e.g., the BBC) and a set of \users, and no other trusted/semi-trusted authority: 

\begin{enumerate}
\itemsep0em
\item{\Users}, possibly organized in groups, compute their (secret) blinding factors, based on the public keys of the other \users, in such a way that they all sum up to zero. They encrypt their local Count-Min Sketch entries (representing their co-view matrix) using these blinding factors, and send the resulting ciphertexts to the \tally. \smallskip

\item The {\tally} receives the encrypted Count-Min Sketch from each \user, aggregates the encrypted counts, and decrypts the aggregates. These are broadcast back to the \users, who use them to recover an estimate of the global similarity matrix and derive personalized ItemKNN-based recommendations.

\end{enumerate}

\descr{Notation.} %
In the rest of this section, we denote with $N$ the number of \users, with $M$ the total number of items, %
and with $L=d \cdot w$ the number of items in a Count-Min Sketch table.
Also, let $\mathbb{G}$ be a cyclic group of prime order $q$ for which the Computational Diffie-Hellman problem (CDH) is hard and $g$ be the generator of the same group. 
$\textrm{H}:\lbrace{0,1\rbrace}^* \rightarrow \Zq$ denotes a cryptographic hash function mapping strings of arbitrary length to integers in $\Zq$. Finally, ``$||$'' denotes the concatenation operator and $a \in_r A$ means that $a$ is sampled at random from $A$.
We assume the system runs on input public parameters $\mathbb{G}, g, q$, where $g$ generates a group of order $q$ in $\mathbb{G}$. 

\begin{figure*}[ht!]
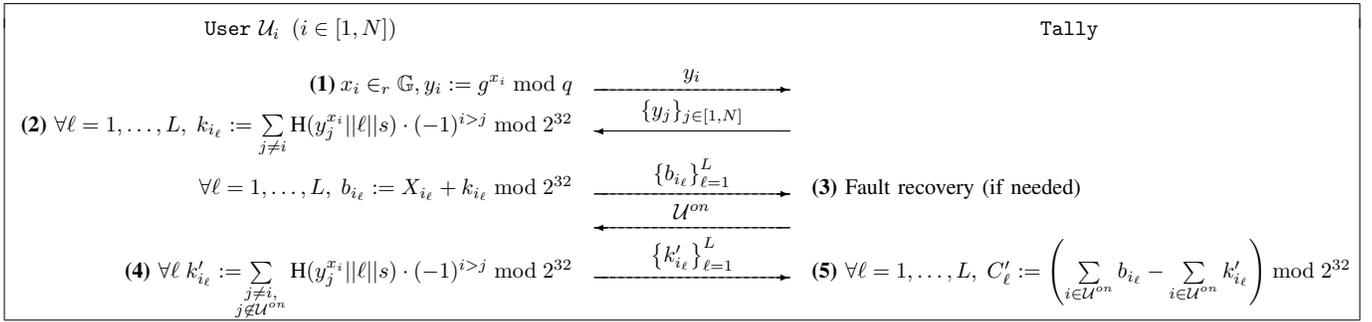

\centering
\resizebox{0.99\linewidth}{!}{
\begin{tabular}{|c|}
\hline\\[-1ex] 
\begin{protocol}{2}
	\participants{\User $\Ui~\left(i\in[1,N]\right)$}{\Tally}\\
		 \textbf{(1)}\: x_i \in_r \mathbb{G}, y_i := g^{x_i} \bmod q & \sends{y_i } &   \\
		\hspace*{-0.25cm}\textbf{(2)}~\forall \ell=1,\ldots,L,\; k_{i_\ell}:=\sum\limits_{\substack{j \neq i}} \textrm{H}(y_{j}^{x_i} || \ell || s) \cdot (-1)^{i > j} \bmod 2^{32} & \receives{\{y_j\}_{j \in[1,N]}} & \\
		\forall \ell=1,\ldots,L,\; b_{i_\ell}:=X_{i_\ell} + k_{i_\ell} \bmod 2^{32}   & \sends{ \left\lbrace b_{i_\ell} \right\rbrace_{\ell = 1}^L} & \mbox{\textbf{(3)}~Fault recovery (if needed)}\\ 
	& \receives{\Uon}  & \\ %
        \textbf{(4)}~\forall \ell\; k'_{i_\ell}:=\hspace*{-0.2cm}\sum\limits_{\substack{j \neq i, \\j  \not\in \Uon }} \textrm{H}(y_{j}^{x_i} || \ell || s) \cdot (-1)^{i > j} \bmod 2^{32} & \sends{ \left\lbrace k'_{i_\ell} \right\rbrace_{\ell = 1}^L} & \textbf{(5)}~\forall \ell=1,\ldots,L,\; C'_{\ell} := \left( \sum\limits_{\substack{i \in \Uon }}b_{i_\ell} - \sum\limits_{\substack{i \in \Uon }}k'_{i_\ell} \right) \bmod 2^{32}\hspace*{-0.25cm}
\end{protocol}
\\ \hline
\end{tabular}
}
\vspace{-0.15cm}
\caption{Cryptographic layer of our private recommender system for online streaming services. At setup \textbf{(1)}, \users compute their secret share and send their public key to the \tally, who broadcasts them to the other \users. 
During the encryption phase \textbf{(2)}, each \user computes the blinding factors, encrypts their Count-Min Sketch and sends it to the \tally. In case not all \users have sent the data, the \tally broadcasts $\Uon$, the subset of \users that did \textbf{(3)}. These compute new blinding factors and send them to the \tally \textbf{(4)}. Aggregate sketches are then recovered by the \tally \textbf{(5)}.}
\vspace{-0.4cm}
\label{fig:press}
\end{figure*}

\subsection{Protocol}
\label{sec:prot}

We now present the details of our proposed protocol. Its cryptographic layer is also summarized in Figure~\ref{fig:press}. 

\descr{Setup.} Each \user $\Ui$ ($i\in[1,N]$) generates a private key $x_i \in_r \mathbb{G}$, and computes and publishes public key $y_i = g^{x_i} \bmod q$. Public keys of all \users are distributed to each other, using a public bulletin board or the \tally itself.

As discussed later in this section, \users might be organized in {\em groups} in order to facilitate aggregation. To ease presentation, we discuss the protocol steps for a single group of \users, as combining aggregates from different groups is trivial and can be done, in the clear, by the \tally.

\descr{Count-Min Sketch construction.} We assume each \user $\Ui$ holds an input vector of data points $I = \lbrace{ I_c \in \mathbb{N},\: c=1,\dots,T \rbrace}$, which represents $\Ui$'s co-view matrix (i.e., $T=M\cdot M/2$). First, $\Ui$ initializes a Count-Min Sketch table $X_i$ %
with all zero entries. In the following, we represent $\Ui$'s Count-Min Sketch table $X_i \in \mathbb{N}^{d \times w}$ as a vector of length $L=d \cdot w$. Then, $\Ui$ encodes $I$ in the Count-Min Sketch using the update procedure described in Section~\ref{sec:cms}, where the following pairwise-independent hash function is employed:\vspace{-0.05cm} 
\[
h(x) = ((ax + b)\bmod p)\bmod w \vspace{-0.05cm}
\]
for $a\neq 0,b$ random integers modulo a random prime $p$. 
At the end of this step, $\Ui$ has built a Count-Min Sketch table $X_i=\{X_{i_\ell}\}_{\ell=1}^L$ (with $L=d\cdot w$ as per Definition~\ref{def:count}).

\descr{Encryption.} To participate in the privacy-preserving sketch aggregation, each \user $\Ui$ first needs to generate blinding factors. At round $s$, for each $\ell=1,\dots,L$, \user $\Ui$ computes:\vspace{-0.2cm}
$$k_{i_\ell} = \sum_{\substack{j=1\\ j \neq i}}^{N} \textrm{H}(y_{j}^{x_i} || \ell || s) \cdot (-1)^{i > j} \bmod q $$ 
where
\vspace{-0.2cm}
\begin{equation*}
(-1)^{i > j} := \begin{cases}
				-1 & \text{if } i > j\\
				1 & \text{otherwise}
				\end{cases}\vspace{0.1cm}
\end{equation*}
Note that the sum of all $k_{i_\ell}$'s equals to zero:\vspace{-0.15cm}
$$\sum_{i=1}^{N}k_{i_\ell} = \sum_{i=1}^{N}\sum_{\substack{j=1\\ j \neq i}}^{N} \textrm{H}(y_{j}^{x_i} || \ell || s)  \cdot (-1)^{i > j} = 0 \vspace{-0.15cm}$$
Then, for each entry $X_{i_\ell}$, $\Ui$ encrypts $X_{i_\ell}$ as $b_{i_\ell} = X_{i_\ell} + k_{i_\ell} \bmod 2^{32}$, as only 32 bits of $b_{i_\ell}$ are enough for our application, and sends the resulting ciphertext to the \tally.

\descr{Aggregation.}
The \tally receives the ciphertexts from the $N$ \users and (obliviously) aggregates the sketches. Specifically, for $\ell=1,\dots,L$, it computes: %
$$C_{\ell} = \sum_{i=1}^{N}b_{i_\ell} = \sum_{i=1}^{N}k_{i_\ell} + \sum_{i=1}^{N}X_{i_\ell} = \sum_{i=1}^{N}X_{i_\ell} \bmod 2^{32}$$
where $C_{\ell}$ denotes the $\ell$-th item in the aggregate Count-Min Sketch table.
$\left\{ C_{\ell} \right\}_{\ell=1}^{L}$, are broadcast back to the \users
(but can obviously be used locally at the \tally too), who use them to recover an estimate of the global matrix and derive personalized recommendations via the ItemKNN algorithm.

\descr{Fault Tolerance.}
If, during the aggregation phase, only a subset of \users report their values $b_{i_\ell}$ to the \tally, the sum of the $k_{i_\ell}$'s is no longer equal to zero and the aggregate items $C_{\ell}$ cannot be decrypted. However, it is possible to recover as follows:
Let $\Uon$ denote the list of \users who have submitted the data in the aggregation phase. The \tally sends $\Uon$ to each $\Ui \in \Uon$.
Then, $\Ui$ computes, for each $\ell=1,\dots,L$,
$$k'_{i_\ell} = \sum_{\substack{j=1\\ j \neq i, j \not\in \Uon }}^{N} \textrm{H}(y_{j}^{x_i} || \ell || s) \cdot (-1)^{i > j} \bmod q $$ 
and sends these values back to the \tally.

Assuming all \users in $\Uon$ submit the values $k'_{i_\ell}$, the \tally can recover the entries in the aggregate sketches (for \users in $\Uon$) by computing:
$$C'_{\ell} = \left( \sum_{\substack{i \in \Uon }}b_{i_\ell} - \sum_{\substack{i \in \Uon }}k'_{i_\ell} \right) \bmod 2^{32}\vspace{-0.2cm}$$

\descr{Groups.} Although the protocol can cope with faults, 
we should nonetheless minimize the probability of missed contributions. Moreover, as discussed in Section~\ref{sec:performance}, the protocol's complexity also depends on the number of \users and, in the case of iPlayer, there can be peaks of hundreds of thousands of users per hour\footnote{\url{http://downloads.bbc.co.uk/mediacentre/iplayer/iplayer-performance-may15.pdf}}. Consequently, we need to organize \users into reasonably sized groups. 
As mentioned earlier, combining aggregates from different groups is straightforward and can be done, in the clear, by the \tally.

We argue that a good choice is between 100 and 1,000 \users per group, as also supported by our empirical evaluation presented later.
There could be a few different ways to form groups: for instance, the \tally could group \users in physical proximity and/or select \users that are  watching/listening a video with at least a couple of minutes left to watch.
Also note that \users not involved in the protocol (or having limited ``history'') can get recommendations too as the \tally can still provide them with the global co-view matrix, which, even though it does not include their own contribution, can be used by the ItemKNN algorithm to derive recommendations.

\descr{Security Analysis.}
The security of our scheme, in the honest-but-curious model, is straightforwardly guaranteed by that of the ``low overhead'' private aggregation scheme by Kursawe et al.~\cite{Kursawe:2011}, which is secure under the CDH assumption.
We modify it
to cope with users faults and to aggregate Count-Min Sketch entries, rather than the actual data, and this does not affect the privacy properties of the scheme.
In case of passive collusions between \users, the confidentiality of the data provided by the non-colluding \users is still preserved.
Finally, note that malicious active \users could report fake values in order to invalidate the final aggregation values, however, %
protocol's integrity could be preserved using verifiable tools such as zero-knowledge proofs and commitments, an extension we leave as part of future work,
along with considering a malicious \tally.

\begin{figure*}[t]
\centering
    \begin{subfigure}[t]{0.4\textwidth}
        \centering
		\includegraphics[width=0.99\columnwidth]{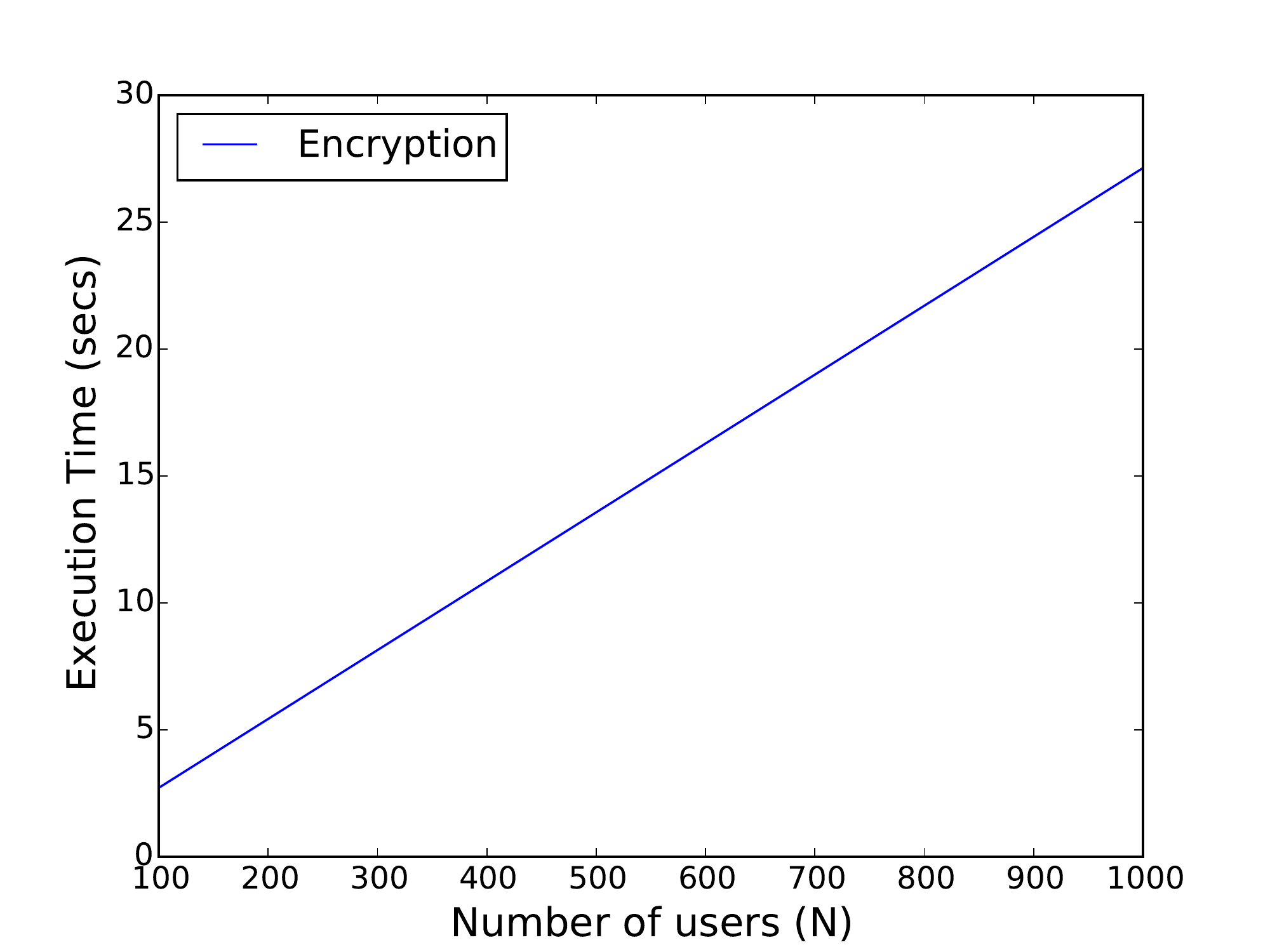}
        \caption{\label{fig:time_client_users} Client}
    \end{subfigure} 
~
    \begin{subfigure}[t]{0.4\textwidth}
        \centering
		\includegraphics[width=0.99\columnwidth]{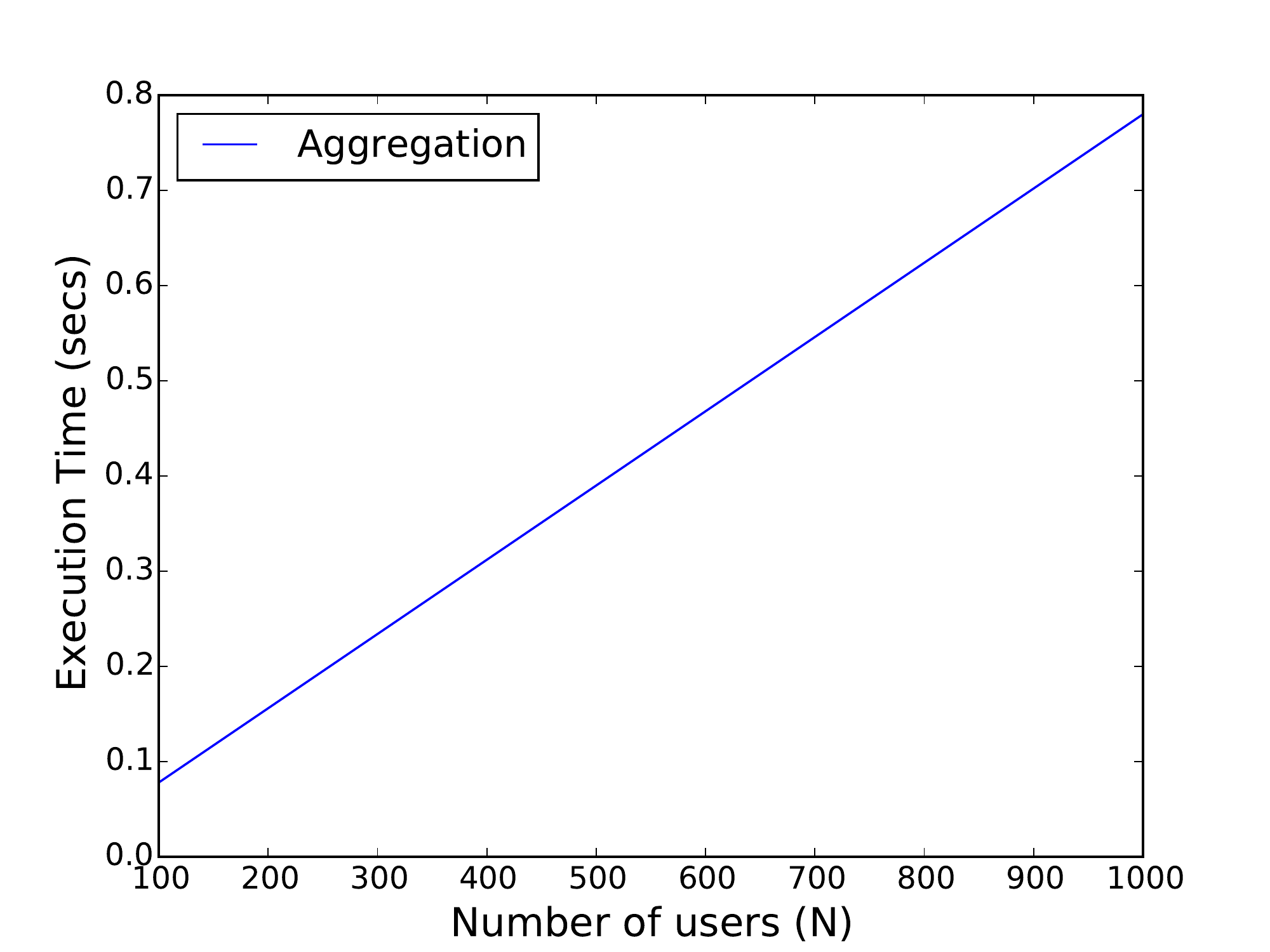}
        \caption{\label{fig:time_server_users} Server}
    \end{subfigure}%
    \vspace{-0.3cm}
\caption{\label{fig:time_users} Execution time  for increasing number of \users (with 700 programs).}
    \vspace{-0.4cm}
\end{figure*}

\subsection{Prototype Implementation}
\label{sec:implementation}
We have implemented the \tally's functionalities as a web application running on the server-side JavaScript environment \emph{Node.js} (or {\em Node} for short).\footnote{\url{https://nodejs.org/}} %
We also use \emph{Express.js}\footnote{\url{http://expressjs.com/}} 
to organize our application into a Model View Controller (MVC) web architecture and \emph{Socket.io}\footnote{\url{http://socket.io/}} to set up bidirectional web-socket connections.
Integrating our solution is as simple as installing a {\em Node} module through the Node Package Manager (NPM) %
and importing it from any web application,
thus requiring no familiarity with the inner workings of the cryptographic and aggregation layers. 
The module for \user's functionalities is modeled as the client-side of the web application and can be run as simple JavaScript code embedded on a HTML page. Therefore, it requires no deployment or installation of any additional software by the \users, but runs directly in the browser, transparently, when \users visit \tally's website.
Our JavaScript implementation is also compatible with smartphone browsers (e.g., the Android version of Chrome), nevertheless, we have also implemented a stand-alone Android application using Apache Cordova.\footnote{\url{https://cordova.apache.org/}} 
The source code of both our browser and Android app \shortVer{will be made available along with the final version of the paper,} \longVer{is available upon request,} so that developers can simply import and extend our code for their own applications.

\descr{Cryptographic Operations.} The cryptographic layer of the protocol is also written in JavaScript, using the Ed25519 curve~\cite{bernstein2011high} implementation available from \emph{Elliptic.js},\footnote{\url{https://github.com/indutny/elliptic}} which supports 256-bit points and provides security comparable to a 128-bit security parameter.
SHA-256 is used for (cryptographic) hashing operations. %

\subsection{Performance Evaluation}
\label{sec:performance}
We now analyze the performance of our system, both analytically (reporting asymptotic complexities) and empirically. %

\descr{Asymptotic Complexities.}
The setup phase carried out by the \user requires $\BigO{N}$ random group points  (where $N$ is the number of total \users) and $\BigO{N}$ messages need to be sent for all the \users to distribute the public keys.
To generate the blinding factors, each \user then needs to perform $\BigO{N}$ exponentiations in $\mathbb{G}$ 
and $\BigO{L \cdot N}$ hashing operations.
Count-Min Sketch encryption (at \user's side) requires $\BigO{L}$ integer additions in $\Zq$, one for each of the $L=\BigO{\log(M^2)}$ Count-Min Sketch entries, while communication complexity amounts to \BigO{L} 32-bits integers for each \user. %
To complete the aggregation, the \tally computes $\BigO{L \cdot N}$ linear operations. 

The use of the Count-Min Sketch significantly speeds up the efficiency of the system. In fact, without them, each \user would need to perform $\BigO{N(M^2)}$ hashing operations and send $\BigO{M^2}$ 32-bit integers, while the \tally would need to compute $\BigO{N(M^2)}$ operations.

\begin{figure*}[t]
\centering
    \begin{subfigure}[t]{0.4\textwidth}
        \centering
		\includegraphics[width=0.99\columnwidth]{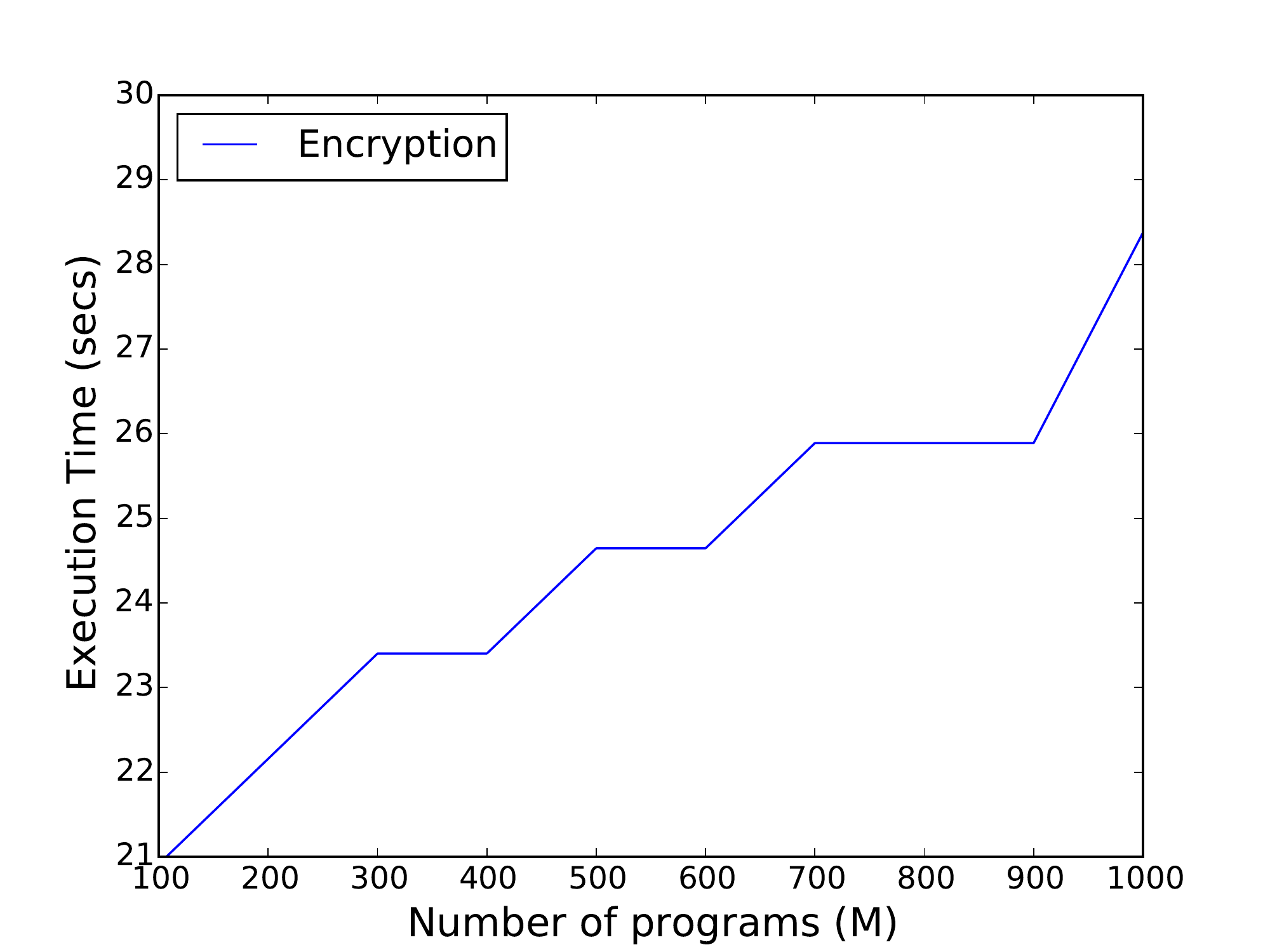}
        \caption{\label{fig:time_client_items} Client}
    \end{subfigure} 
~
    \begin{subfigure}[t]{0.4\textwidth}
        \centering
		\includegraphics[width=0.99\columnwidth]{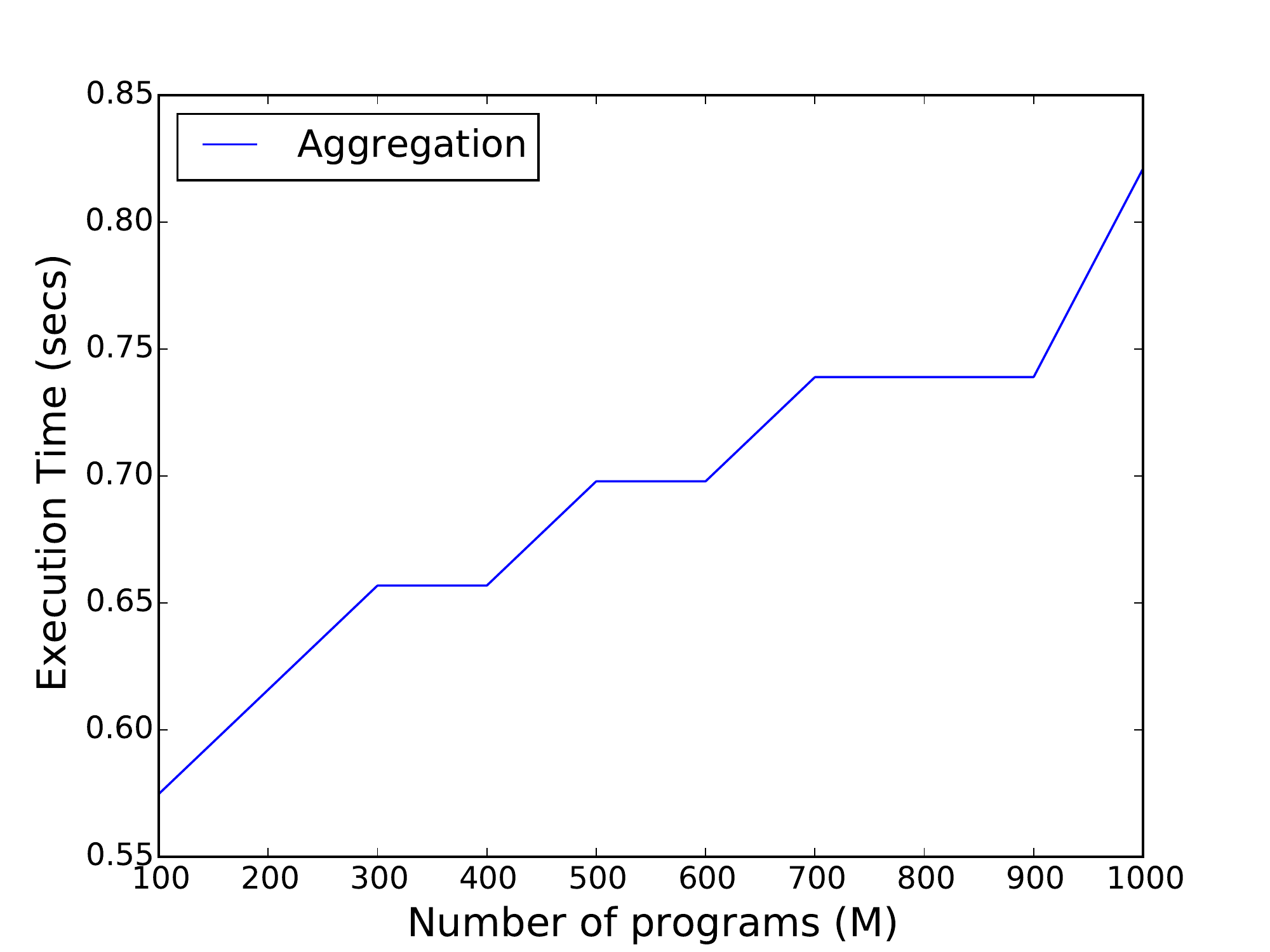}
        \caption{\label{fig:time_server_items} Server}
    \end{subfigure}%
    \vspace{-0.3cm}
\caption{\label{fig:time_items} Execution time  for increasing number of programs (with 1,000 \users).}
\vspace{-0.5cm}
\end{figure*}

\begin{figure*}[t]
\centering
    \begin{subfigure}[t]{0.4\textwidth}
        \centering
		\includegraphics[width=0.99\columnwidth]{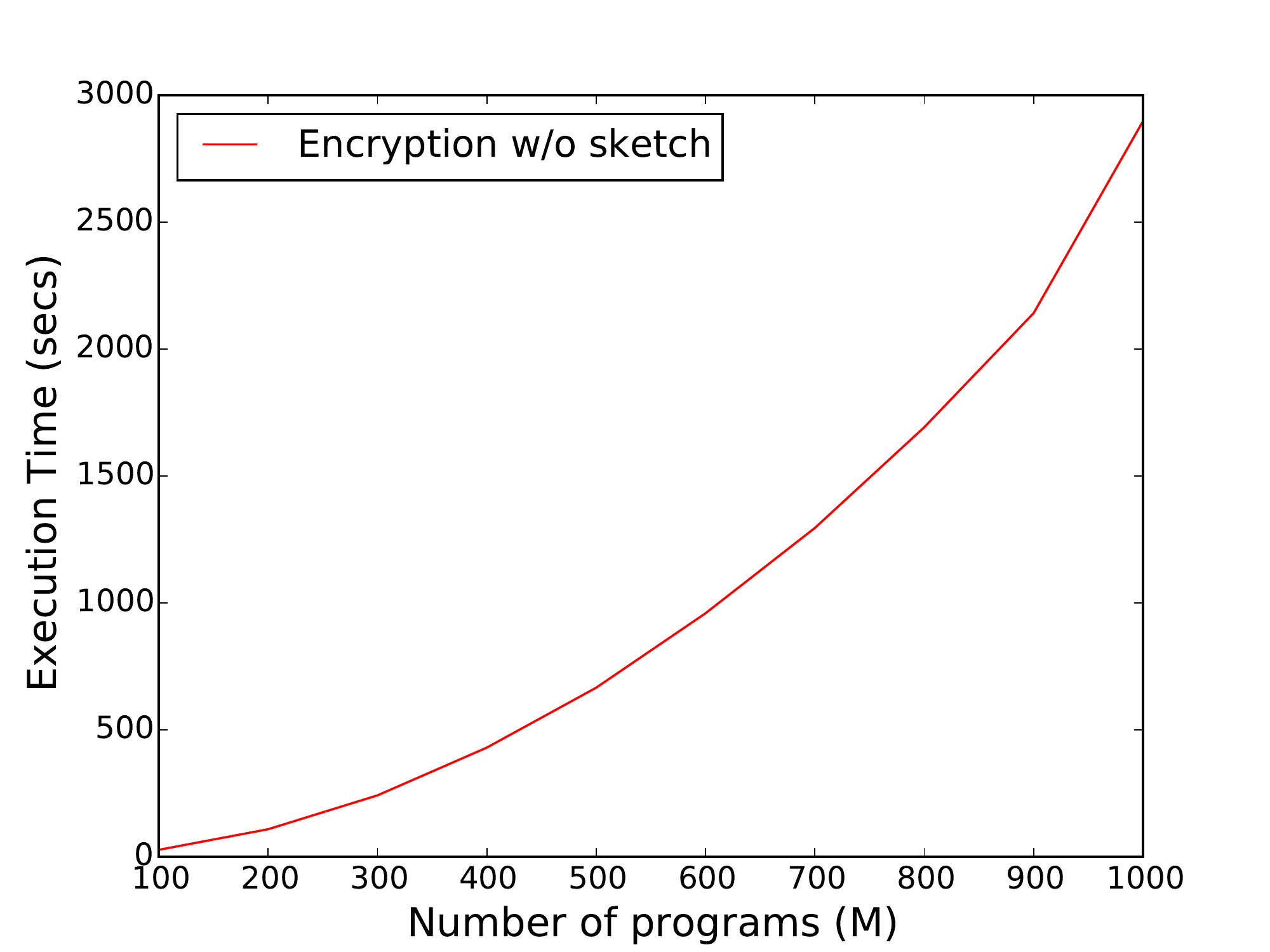}
        \caption{\label{fig:wo_time_client_items} Client}
    \end{subfigure} 
~
    \begin{subfigure}[t]{0.4\textwidth}
        \centering
		\includegraphics[width=0.99\columnwidth]{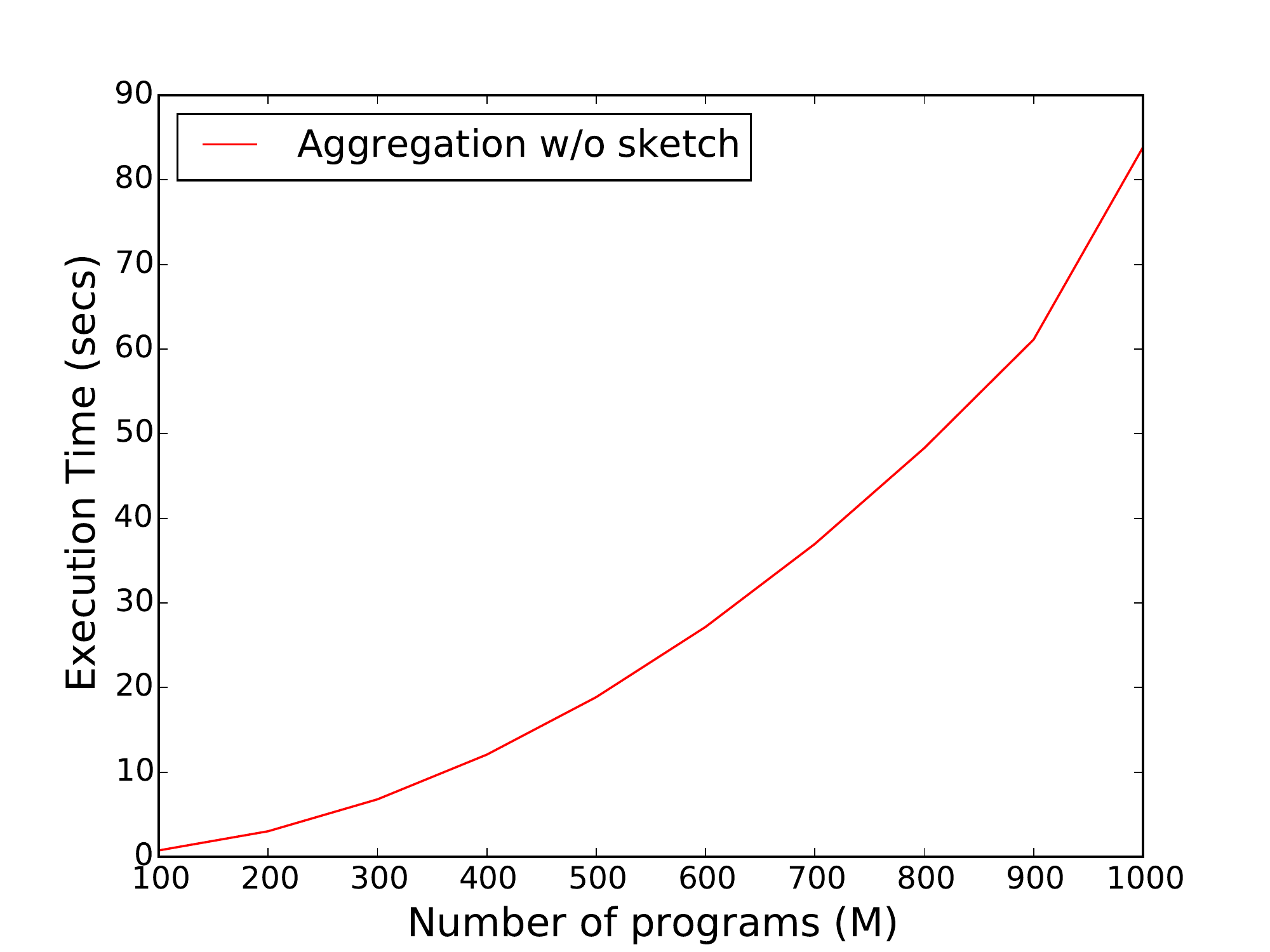}
        \caption{\label{fig:wo_time_server_items} Server}
    \end{subfigure}%
    \vspace{-0.3cm}
\caption{\label{fig:wo_time_items} Execution time  for increasing number of programs (with 1,000 \users)  {\em without} Count-Min Sketch.}
\vspace{-0.2cm}
\end{figure*}

\descr{Computation Overhead.} 
We have also simulated the execution of our private recommender system and measured execution times (averaged over 100 iterations) for all operations. 
Simulations have been performed on a machine running Ubuntu Trusty (Ubuntu 14.04.2 LTS), equipped with a 2.4 GHz CPU i5-520M and 4GB RAM. %

In Figure~\ref{fig:time_users}, we plot running times of protocol's client- and server-side for an increasing number of \users, fixing the number of programs to 700 (the average number of programs available on iPlayer) and the sketch parameters to $\epsilon = \delta = 0.01$ (see Definition~\ref{def:count}). 
Using this setting, the number of rows $d$ and columns $w$ of the Count-Min Sketch amounts to  
$
d=18 \:, \:w= 272
$ 
leading to a Count-Min Sketch of size $L = d \cdot w = 18 \cdot 272 =$ 4,896. 
Running times grow linearly in the number of \users. 
As illustrated in Figure~\ref{fig:time_client_users}, the encryption, performed by each \user (see step (2) in Figure~\ref{fig:press}), takes 2.7 seconds with 100 \users and 27 seconds with 1,000 \users, while
Figure~\ref{fig:time_server_users} reveals that \tally completes the aggregation (step (5) in Figure~\ref{fig:press}) in 78ms (resp., 780ms) with 100  (resp., 1,000) \users.

We then measure the execution time for an increasing number of programs and a fixed number of \users, i.e., 1,000.
Figure~\ref{fig:time_client_items} illustrates running times' logarithmic growth for encryption, ranging from 21 seconds with 100 programs to 28 seconds with 1,000 programs. Figure~\ref{fig:time_server_items} illustrates \tally's execution times for the aggregation, which approximately range from 600ms to 800ms.
Note that the ``stair'' effect of the plots in Figure~\ref{fig:time_items} is due to the fact that the Count-Min Sketch size can be the same with close numbers of programs.

Without the compression factor of the Count-Min Sketch, the running times for both \user and \tally would grow linearly in the size of the co-view matrix  (i.e., $M\cdot M/2$), yielding remarkably slower executions. As illustrated in Figure~\ref{fig:wo_time_client_items}, with 1,000 \users and 1,000 programs, running time for each \user amounts to almost 50 minutes instead of 28 seconds using the sketch, whereas, the aggregation at the \tally completes in almost one and a half minute (versus less than one second using Count-Min Sketch). 
Finally, execution time of the \emph{ItemKNN} operations carried out at \user's side, with 700 programs, amounts to 850ms for each \user.

\descr{Communication Overhead.} In Table~\ref{tab:bytes}, we report the amount of bytes exchanged between all parties for different number of \users and Count-Min Sketch sizes, fixing the number of programs to 700.
Note that, without the compressing factor of the sketch, with 700 programs, each \user would have to send 960KB instead of 20KB.

\begin{table}[t]
\centering
\resizebox{0.99\columnwidth}{!}{%
\centering
\begin{tabular}{r|rrr|r}
\#\Users & Bytes  & &  Sketch Size & Bytes\\
 & (\Tally to \User) & & & (\User to \Tally) \\ %
 \cline{1-2} \cline{4-5}
100  & 3,200  &  & 4,896 & 19,584 \\ 
200  & 6,400  &  & 2,448 & 9,792  \\
300  & 9,600  &  & 1,638 & 6,552  \\
400  & 12,800 &  & 1,224 & 4,896  \\ 
500  & 16,000 &  & 972  & 3,888  \\ 
600  & 19,200 &  & 810  & 3,240  \\
700  & 22,400 &  & 702  & 2,808  \\
800  & 25,600 &  & 612  & 2,448  \\
900  & 28,800 &  & 540  & 2,160  \\ 
1000 & 32,000 &  & 486  & 1,944  \\
\end{tabular}
}
\caption{Bytes exchanged by \user and \tally for different \#\users and size of the Count-Min Sketch, considering 700 programs.}\label{tab:bytes}
\vspace{-0.4cm}
\end{table}

\begin{figure*}[t]
\centering
    \begin{subfigure}[t]{0.4\textwidth}
        \centering
		\includegraphics[width=0.94\columnwidth]{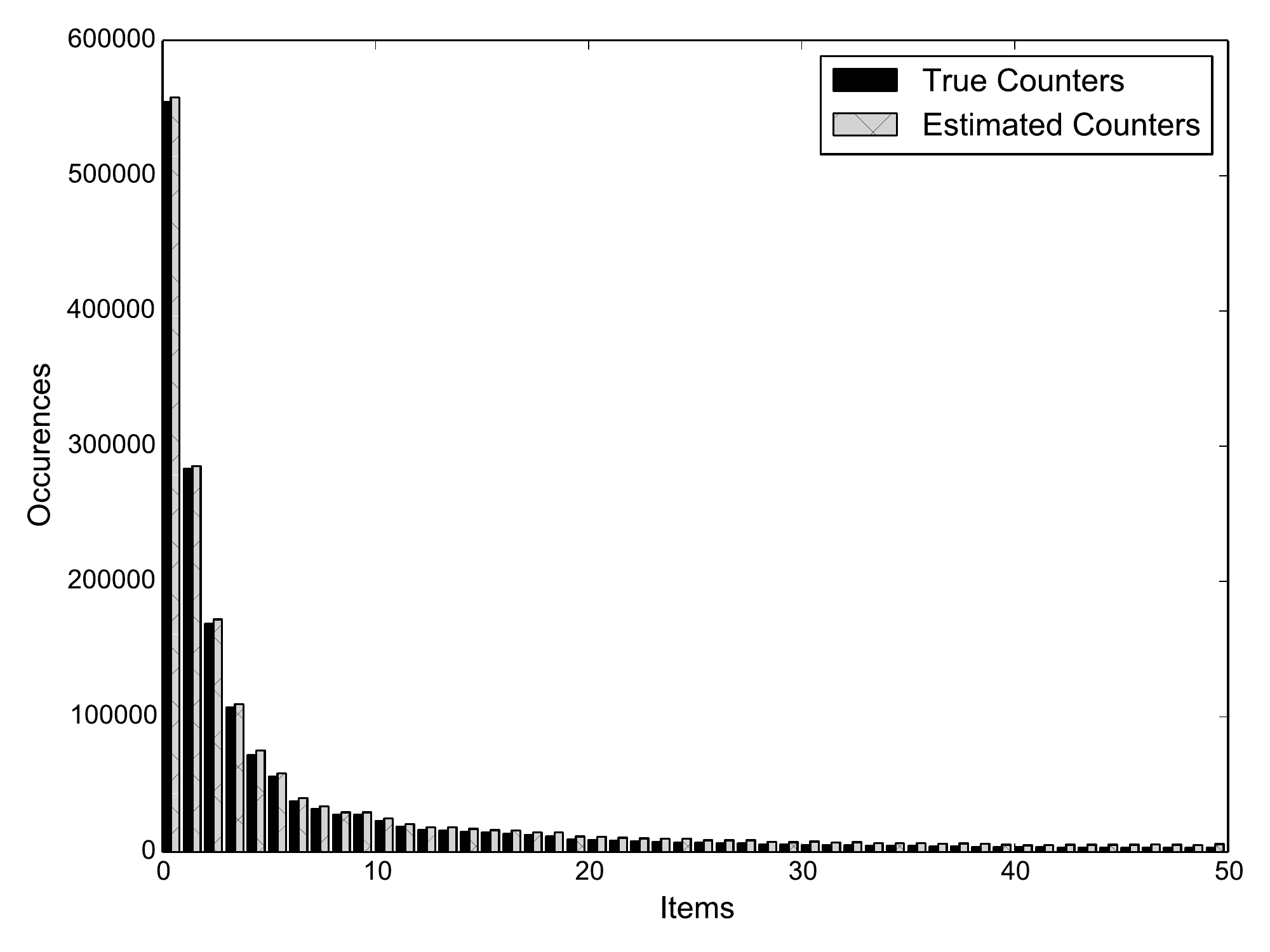}
        \caption{\label{fig:top50} True vs estimated counters}
    \end{subfigure} 
~
    \begin{subfigure}[t]{0.4\textwidth}
        \centering
		\includegraphics[width=0.88\columnwidth]{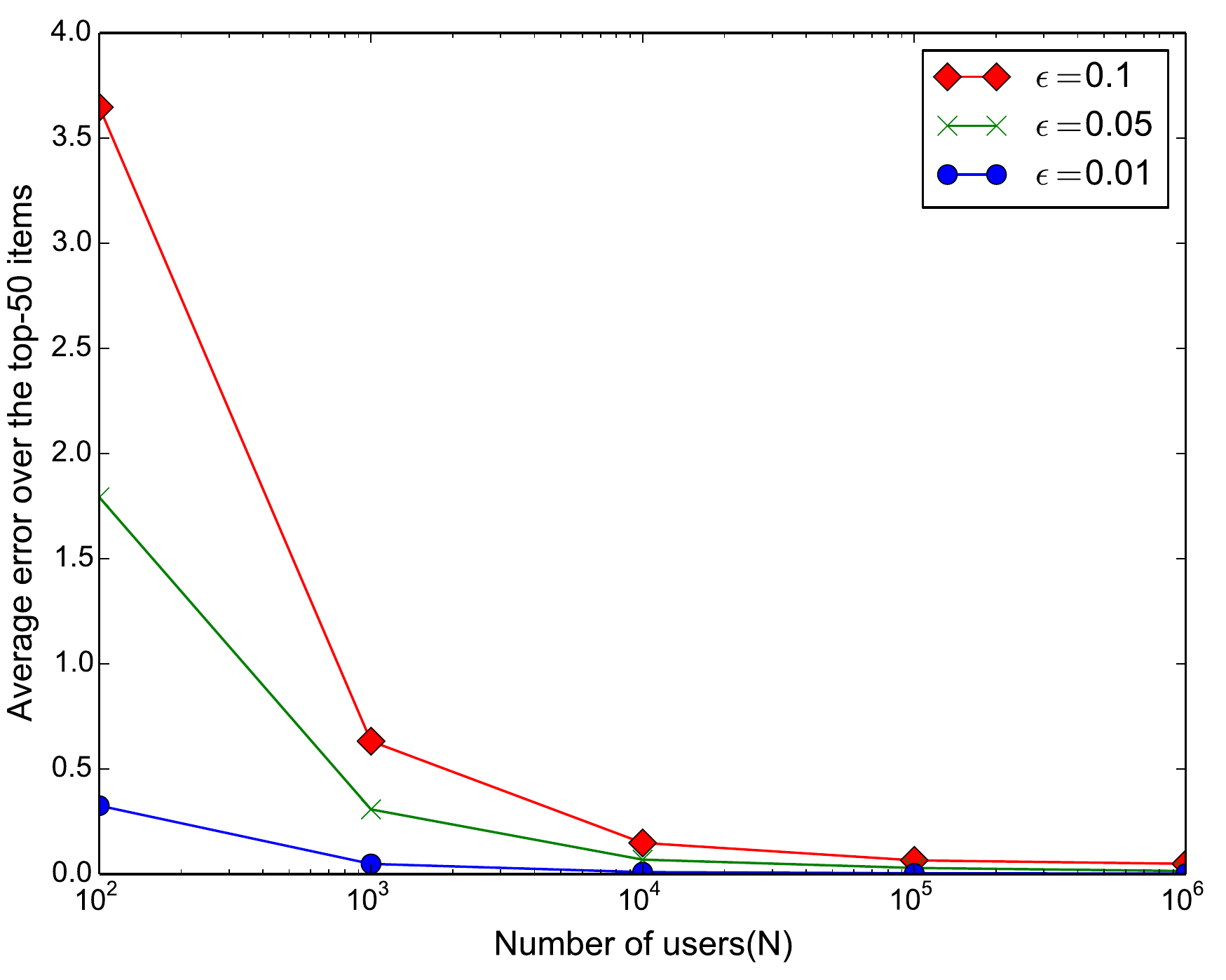}
        \caption{\label{fig:sketch-accuracy} Average error for different values of $\epsilon$}
    \end{subfigure}%
    \vspace{-0.3cm}
\caption{\label{fig:figs} Visualizing the accuracy of the Count-Min Sketch for the most 50 frequent items (with 700 programs and sketch size 4,896).}
    \vspace{-0.3cm}
\end{figure*}

\descr{Accuracy Estimation.} Finally, we evaluate the accuracy loss due to the use of Count-Min Sketch, specifically, over the most 50 frequent items,
using a synthetic dataset sampled from a zipfian distribution simulating a million \users.
We set the Count-Min Sketch parameters to be $\epsilon = 0.01$ and $\delta = 0.01$ as we have measured an acceptable accuracy loss level introduced by the Count-Min Sketch (see below). %
Once again, we fix the number of programs to $M=\text{700}$, leading to a Count-Min Sketch of size $L =$ 4,896.
Figure~\ref{fig:top50} shows that the Count-Min Sketch estimation over the most 50 frequent items is almost indistinguishable from the true population.

We also plot, in Figure~\ref{fig:sketch-accuracy}, the average error, defined as
$\vert \hat{c_i} - c_i \vert/\sum_{j} |c_j|$,
over the most 50 frequent items with an increasing number of \users, while  fixing $M=\text{700}$, $\delta = 0.01$ (yielding a total number of items to update on the Count-Min Sketch of $T=M\cdot M/2=$~245,000) and three choices  of the $\epsilon$ parameter, i.e., $0.01,0.05$, and $0.1$.
The average error decreases with more \users and smaller values of $\epsilon$. Standard deviation values are infinitesimal, thus, we do not include them in the plot as they would not be visible.

\section{Private Aggregate Location Prediction }
\label{sec:smartphone}
The rapid proliferation of smartphones, with 2 billion estimated users by the year 2016~\cite{telegraph}, makes it increasingly easy (and appealing) to track users' locations and movements using sensors like GPS and WiFi.
This contextual information can be extremely useful to train machine learning algorithms and predict future events, paving the way for anticipatory mobile computing~\cite{pejovic2013anticipatory}.
Location and movement models can be used, e.g., for traffic mitigation, road  monitoring, and hazard detection~\cite{cartel}, as well as to guide decision frameworks to respond to anomalies and disruptions on short notice.

Pervasive location sensing, however, raises important privacy concerns as single individuals' movements can easily be tracked and sensitive information could be exposed. If home and work locations can be deduced from anonymized location traces, single individuals can be uniquely re-identified~\cite{golle2009anonymity}.
Moreover, location patterns have been shown to leak personal information, e.g., taxi drivers' religion and individuals' visits to gentleman's clubs.\footnote{See \url{http://on.mash.to/1ByncHD} and \url{https://goo.gl/Ta5JYG}.} 

In this section, we instantiate a smartphone application enabling \users to report, to a service provider (\tally), their locations over time. \Users' privacy is protected as only aggregate (over many \users) location statistics are disclosed. We then show how these statistics can be used to train a model and predict future movements, and support private computation and prediction of ``heat maps'' relying on the aggregate counts of people in a given area over a period of time. %

\descr{System Model.} We operate in the same model as our privacy-friendly recommender system (cf.~Section~\ref{sec:prot}), involving a \tally that privately aggregates location statistics contributed from a set of \users, and re-use the same cryptographic layer. Once again, we support efficient computation of private statistics using (i) Count-Min Sketch's succinct data representation and (ii) privacy-preserving aggregation with \users' blinding factors summing up to zero.

\descr{Overview.} We assume a 2-D space territory $\mathcal{R}$ is partitioned into a grid of $|S| = p \times p$ cells ($S = \lbrace{ S[1,1], S[1,2], \dots, S[p,p]  \rbrace}$), and $t$ finite intervals (time slots) $\left[ t_{j-1}, t_{j} \right]$, where $j \in \mathbb{N}^+$.
Let $S_i^{(t_j)}$ be the grid containing, for each cell, the number of times the \user $\Ui$ has logged her position (using a GPS measurement) within that particular cell over $t \in \left[ t_{j-1}, t_{j} \right] $.
\User $\Ui$, for each time slot $\left[ t_{j-1}, t_{j} \right]$, builds the grid $S_i^{(t_j)}$ with locations logged over time, maps the grid into a Count-Min Sketch, and sends the encrypted sketch to the \tally. 
This aggregates and decrypts them, reconstructing the grid containing the (estimated) aggregate locations. 

The location statistics can be used to display `heat maps'' (e.g., a graphical representation of congestion), or to perform time-series based prediction over a sequence of heat maps.
Using an Exponential Weighted Moving Average (EWMA) model (see Section~\ref{sec:timePred}), we can predict the future popularity of a cell, by relying on the past (approximated) observations for that cell.
Other machine learning techniques, e.g., Multivariate Support Vector Machines or Logistic Regression, could also be used for the prediction, but we consider it to be beyond the scope of this paper to investigate new predictors.

\descr{The San Francisco Cabs Dataset.}
To evaluate the feasibility of our intuition, we use a publicly available dataset containing mobility traces of San Francisco taxi cabs.\footnote{\url{http://cabspotting.org/}}
The dataset contains  11 million GPS coordinates, generated by 536 taxis over almost a month in May 2008. 
\begin{figure}[t]
\centering
\includegraphics[width=0.8\columnwidth]{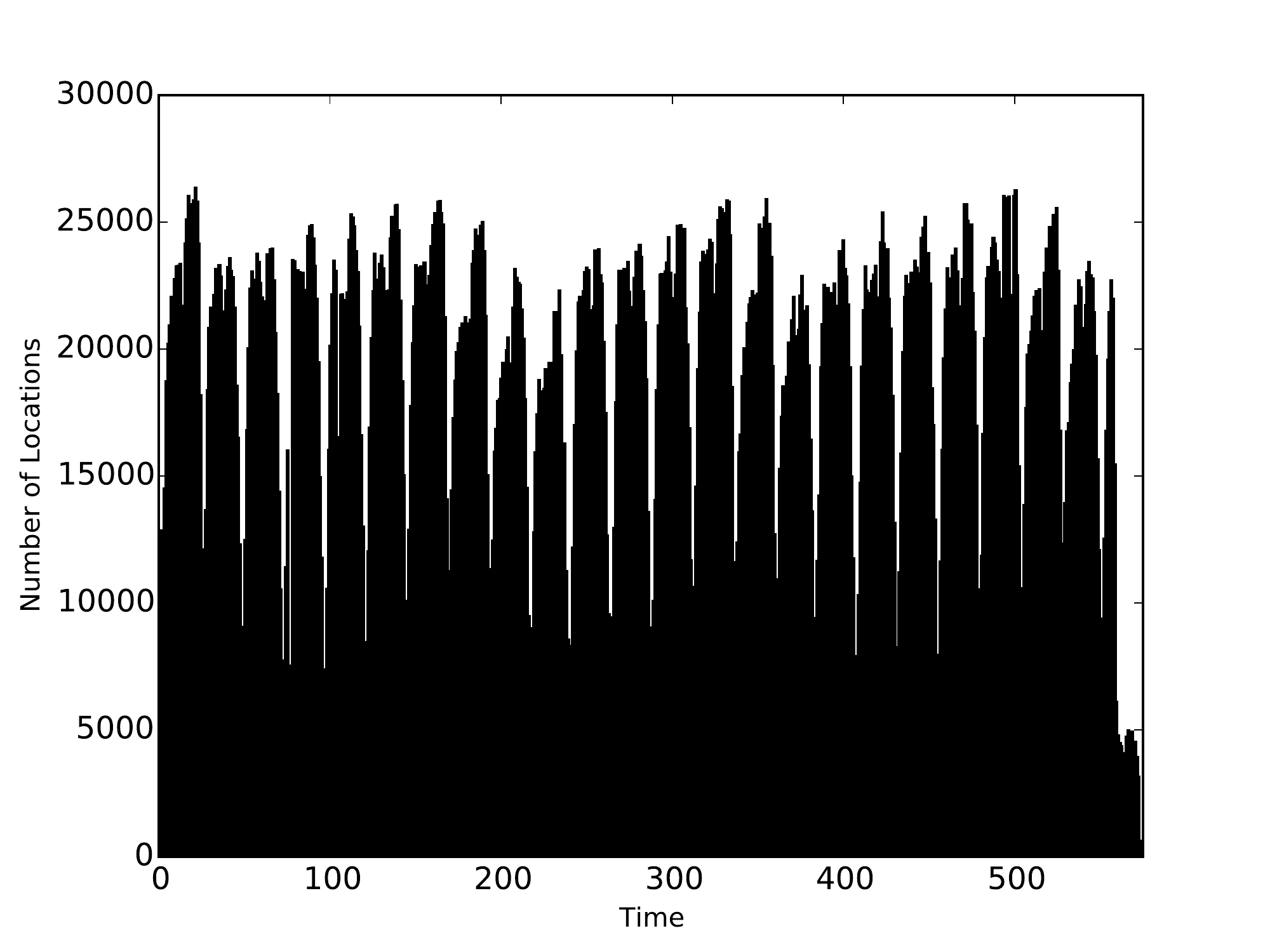}
\vspace{-0.15cm}
\caption{\label{fig:heat-slots} Number of taxi locations over time.}
\vspace{-0.2cm}
\end{figure}
We group the taxi locations in time slots of one hour, leading to a total of $575$ epochs.
Figure~\ref{fig:heat-slots} shows 
the presence of weekly and daily patterns in the number of taxi locations over time (i.e. hourly time slots) and peaks of roughly 25,000 total hourly contributions.
\begin{figure}[t]
\centering
\includegraphics[width=0.81\columnwidth]{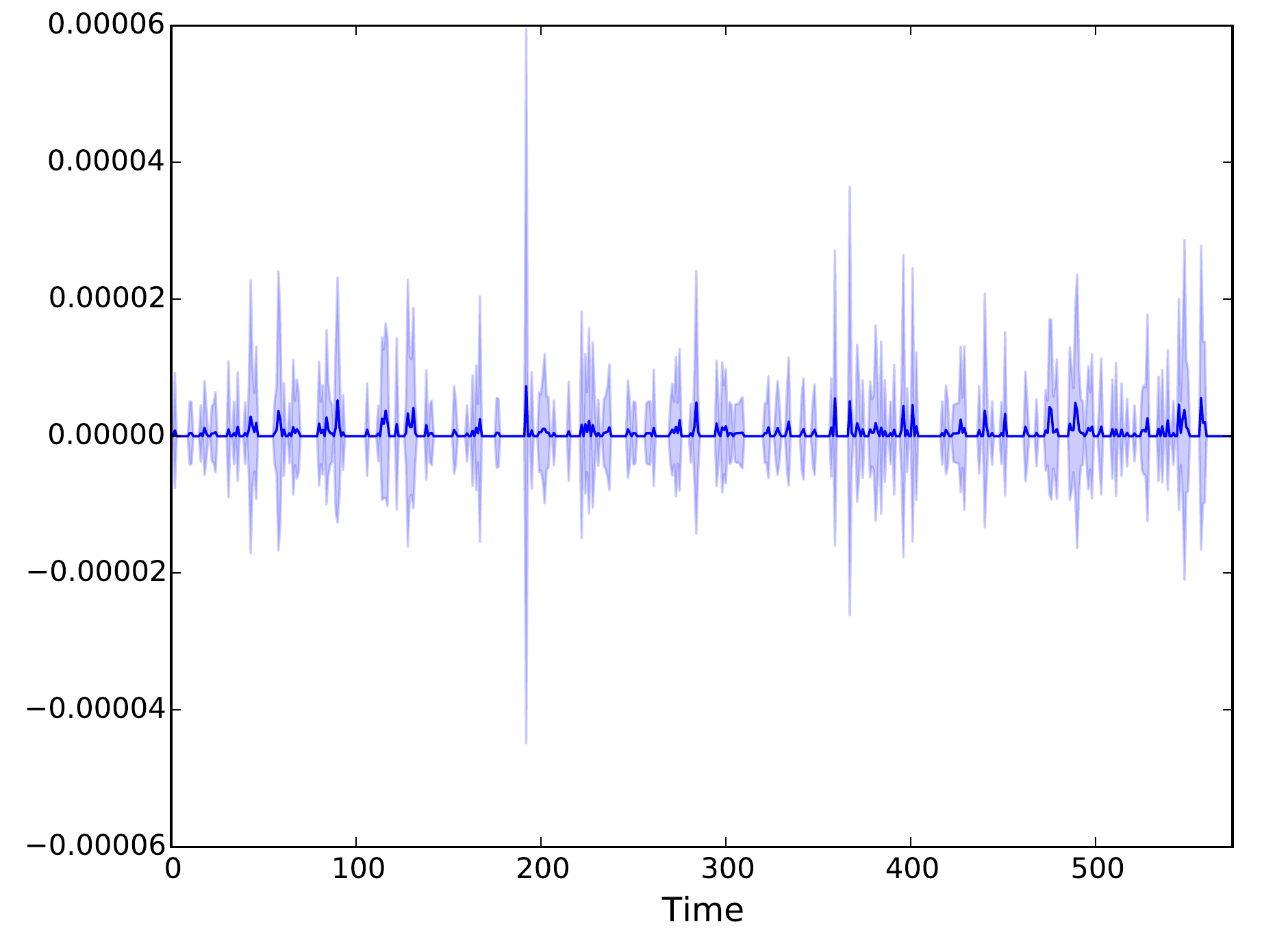}
\vspace{-0.1cm}
\caption{\label{fig:heat-error} Average error introduced by the Count-Min Sketch on the aggregate statistics for the top-100 locations.}
\vspace{-0.25cm}
\end{figure}

\descr{Succinct Data Representation.} We investigate whether succinct data representation could be applied to the problem of collecting location statistics, and measure the accuracy loss introduced by the Count-Min Sketch's compact representation. In Figure~\ref{fig:heat-error}, we plot the average error defined as
$\vert \hat{c_i} - c_i \vert/\sum_{j} |c_j|$
and the relative standard deviation over the most 100 popular cells for each time slot, while fixing $\epsilon = \delta = 0.01$ 
and the total number of cells to $|S| = 100 \times 100$ (yielding a Count-Min Sketch of size $L = 3,808$).
Observe that the average error is infinitesimal for every time slots.

\begin{figure}[t]
\centering
\includegraphics[width=0.82\columnwidth]{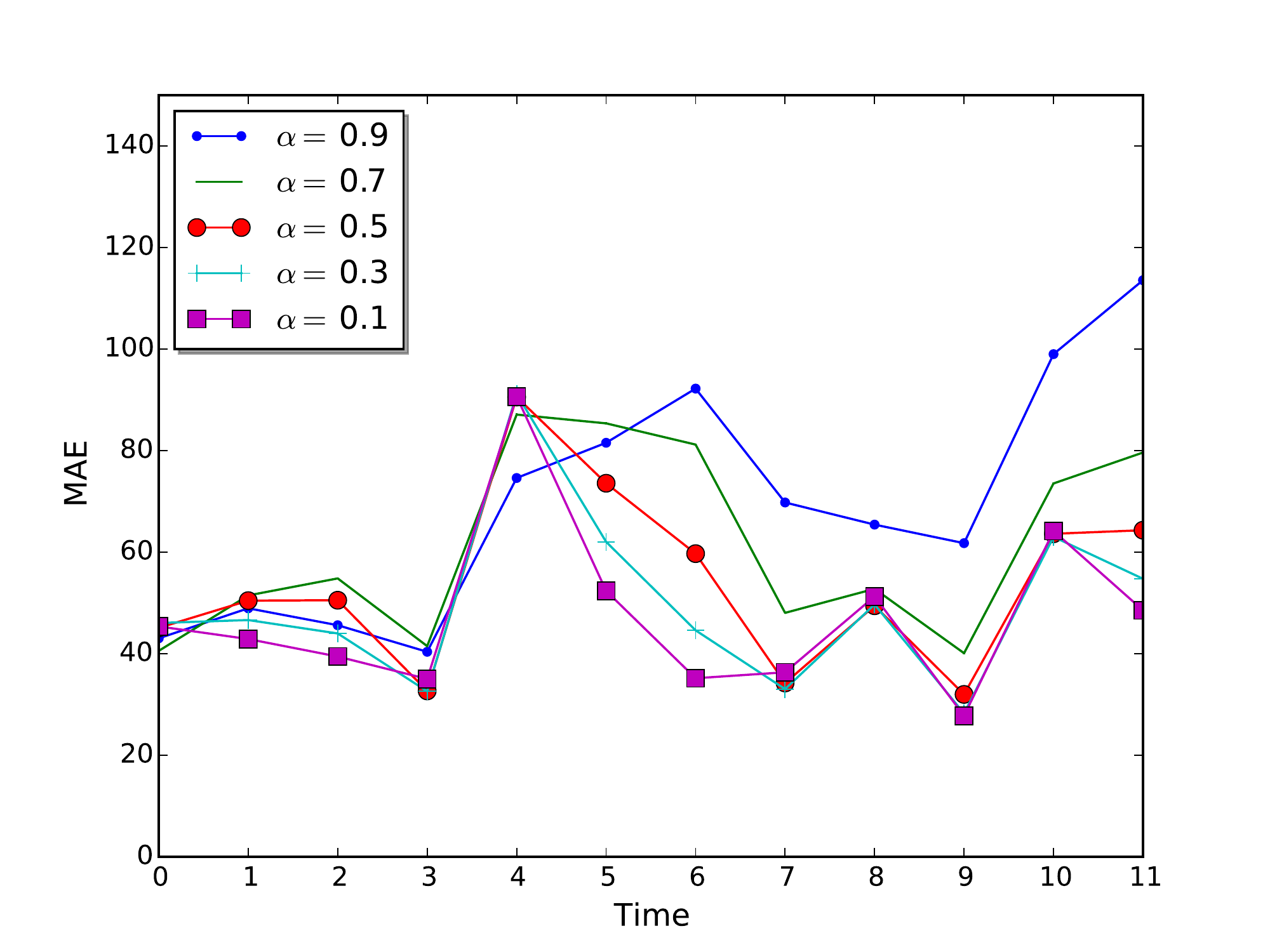}
\vspace{-0.1cm}
\caption{\label{fig:heat-pred} Mean absolute error in the prediction for different values of prediction algorithm's parameter $\alpha$.}
\vspace{-0.15cm}
\end{figure}

\descr{Heat Map Prediction.} Next, we focus on predicting future heat maps using the EWMA algorithm introduced in Section~\ref{sec:timePred}. We start by evaluating the accuracy of EWMA-based prediction relying on the aggregates collected {\em without using the Count-Min Sketch}. We perform the prediction over a subset of $12$ consecutive epochs having the maximum number of reported locations, giving the past 24 hours observations as input to the EWMA algorithm.
Figure~\ref{fig:heat-pred} plots the Mean Absolute Error (MAE) in the prediction compared to the ground truth over the most 100 popular cells, considering different values of 
$\alpha$, i.e., EWMA's smoothing coefficient (cf.~Section~\ref{sec:timePred}).
The plot shows that, in almost all slots, lower values of $\alpha$ lead to more accurate results. 
\begin{figure}[t]
\centering
\includegraphics[width=0.8\columnwidth]{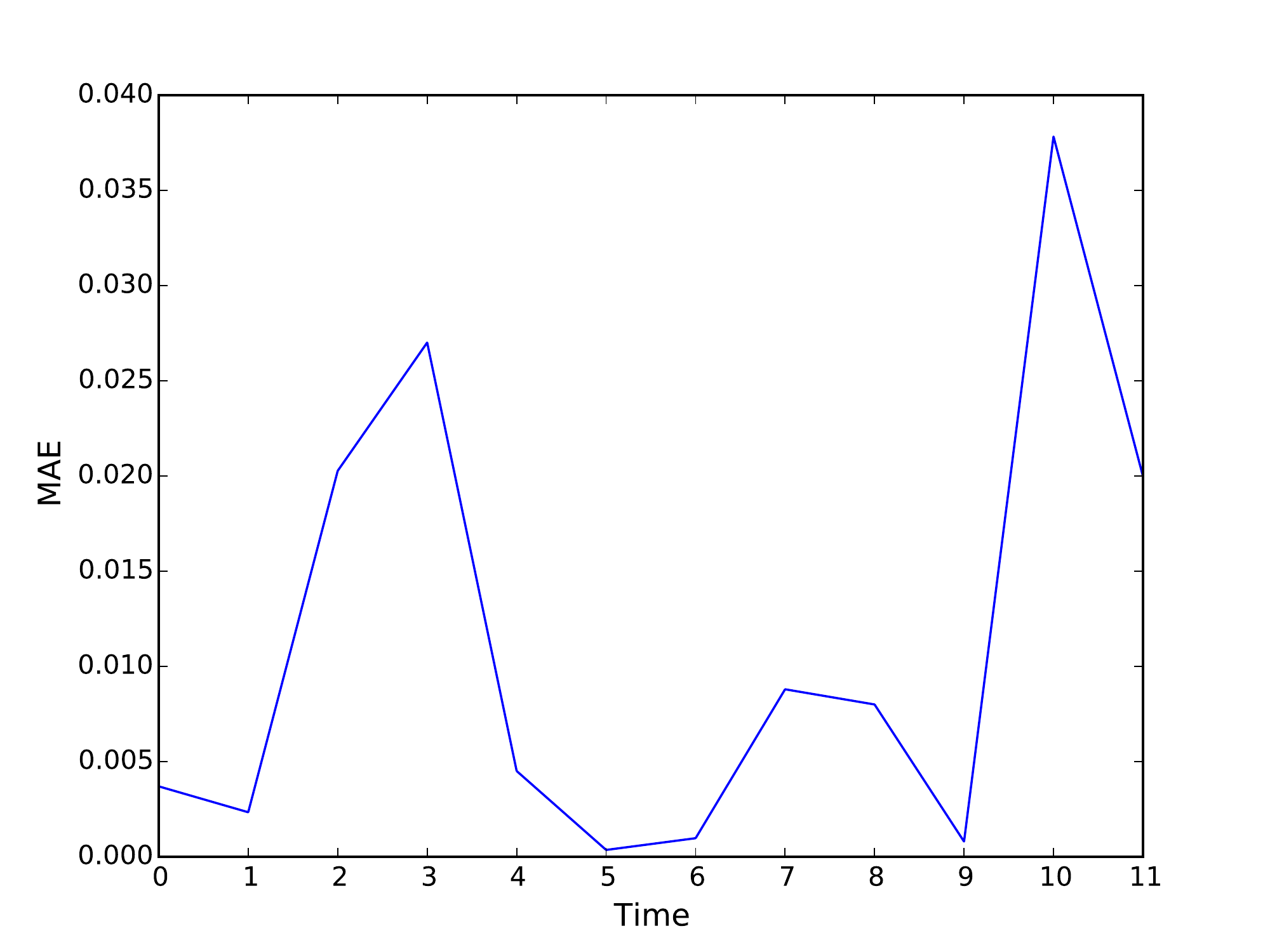}
\vspace{-0.1cm}
\caption{\label{fig:heat-comp} Mean absolute error introduced by the Count-Min Sketch on the prediction accuracy.}
\vspace{-0.45cm}
\end{figure}

We then perform the prediction over the approximate heat maps, i.e., {\em using the sketches}. We focus on the same time slot, and fix $\alpha = 0.1$.
Figure~\ref{fig:heat-comp} shows the error introduced by the Count-Min Sketch in the prediction, for each time slot considered, with respect to the prediction based on the {\em ``real''} heat maps. 
We observe that this error, while fluctuating, is appreciably low for every prediction, thus confirming the feasibility of our techniques for the problem of privately predicting future heat maps.

Once again, we have implemented our techniques in JavaScript, with the server-side running as a {\em Node} module, and client-side running as an open-source Android application built using Apache Cordova.
\shortVer{Source code will be made available along with the final version of the paper.} \longVer{Source code is available upon request.}
Note that, due to space limitations, a performance evaluation of our implementations is not presented in this version as it would anyway mirror the one presented in Section~\ref{sec:service}.

\section{Gathering Statistics on Tor Hidden Services}
\label{sec:Tor}

The privacy-preserving collection of statistics  using efficient data structures, seeking a trade-off between accuracy and efficiency, has also interesting applications in non-user facing settings such as 
collecting network statistics from servers or routers. 
In this section, we present a novel mechanism geared to  privately gather statistics in the context of the Tor anonymity network~\cite{Tor}. The Tor project has recently received funding to improve monitoring of load and usage of Tor hidden services.\footnote{\url{https://www.torproject.org/docs/hidden-services.html.en}} 
This motivates them to extract aggregate statistics about the number of hidden service descriptors from multiple Hidden Service Directory authorities. In order to ensure robustness, the Tor project has determined that the median -- rather than the mean -- of these volumes should be calculated, which is beyond privacy-friendly statistics approaches like Privex~\cite{elahi2014privex}. %

In this section, we first describe the protocol for estimating median statistics using Count Sketch, then, we present the design and deployment of its prototype implementation, along with its performance evaluation.

\subsection{Private Median Estimation using Count Sketch}
We rely on the Count Sketch~\cite{charikar2002finding} data structure, which closely resembles Count-Min Sketch, used in Sections~\ref{sec:service}--\ref{sec:smartphone}. Recall from Section~\ref{sec:cms} that building a Count Sketch follows the same process as a Count-Min Sketch, thus leading to a $d \cdot w$ table of positive integer values, whereas, the estimation of an item's frequency is slightly different: for each row, $d_i$, a hash function is applied to the item leading to a column $w_j$. An unbiased estimator of the frequency of the item is the value at this position minus the value at an adjacent position -- and the median of those estimators is the final estimated frequency. What is key to the success of our techniques is that the estimate of the frequency of specific values, as well as sets of values, is a simple linear sum of Count Sketch entries; computing it does not require non-linear (e.g., \emph{min}) operations as for the Count-Min Sketch. 
For this application, we build on privacy-preserving data aggregation based on threshold public-key encryption, 
specifically, an Additively Homomorphic Elliptic-Curve variant of El Gamal (AH-ECC)~\cite{benaloh1994dense}, summarized below. 
This allows us to seamlessly tolerate missing contributions -- following an approached first proposed by Jawurek et al.~\cite{jawurek2012fault}.

\descr{AH-ECC} consists of the following three algorithms (using a multiplicative notation):
\begin{compactenum}
\item {\em KeyGen}$(1^\tau)$: Given a security parameter $\tau$, choose an elliptic curve $E$ and $(g_1,g_2)$ public generators on $E$, generating a group of order $q$. Choose a random private key $x \in \mathbb{Z}_q$, define the public key as $pk = {g_1}^{x}$, and output public parameters $(E, g_1, g_2, \mathit{pk})$ and private key $x$. 
\item {\em Encrypt}$(m, \mathit{pk})$: The message $m$ is encrypted by computing two elliptic curve points as 
$(A, B) := ({g_1}^r, \mathit{pk}^r {g_2}^m)$, where $r \in \mathbb{Z}_q$ is selected at random. The ciphertext is thus the tuple of points $(A, B)$.
\item {\em Decrypt}$(A, B, x)$: Decryption is performed by computing the element $B A^{-x} = {g_2}^m$. We can achieve constant time decryption by pre-computing a table of discrete logarithms which is then used to recover $m$ from ${g_2}^m$ (this solution is practical  for small values of $m$).
\end{compactenum}
AH-ECC is additively homomorphic since an element-wise multiplication of ciphertexts yields an encryption of their sum.%

\descr{Setup.} %
Our system relies on a set of authorities that can jointly decrypt a ciphertext from the AH-ECC additively homomorphic public-key cryptosystem.
During setup, each authority generates their public and private key and a group public key is computed by multiplying all the authorities' public keys. 
Note that we operate in a distributed system setting (i.e., the Tor network), therefore, similar to PrivEx~\cite{elahi2014privex}, one can easily instantiate decryption authorities.

\descr{Protocol.} Using Count Sketch, we can collect a number of private readings from Hidden Service Directories (HSDir), and compute an approximation of the median. 
Each HSDir builds a Count Sketch, inserts its private values into it, encrypts it, and sends it to the authorities. These aggregate all sketches by homomorphically adding them element-wise, %
yielding an encrypted sketch summarizing the set of all HSDir values. 

Once the authorities have computed the aggregate sketch, an interactive divide-and-conquer algorithm is applied to estimate the median given the range of its possible values is known. At each iteration, the number of sample values in the range is known, starting with the full range and all values received. The range is then halved and the sum of all elements falling in the first half of the range is jointly decrypted. If the median falls within first half of the range it is retained for the next iteration, otherwise the second half of the range is considered at the next iteration. The process stops once the range is a single element. 
Following the master theorem~\cite{cormen2001introduction}, we know that this process converges in $\BigO{\log n}$ steps, for $n$  elements in the domain of the values/median. Due to frequency estimations for the ranges using Count Sketches that provide noisy estimates, we expect this median to be close, but possibly not exactly the same as the true sample median, depending on the Count Sketch parameters $\delta$ and $\epsilon$.

\descr{Output Privacy.} Note that this process is not ``perfectly'' private in a traditional secure computation setting, as the volume of reported values falling within the intermediate ranges considered is leaked. This may be dealt with in two ways: (1) the leakage may be considered acceptable and the algorithm run as described, or (2) the technique can be enhanced to provide differential privacy by adding noise to each intermediate value.

\descr{Differentially Private Estimates.} The sensitivity~\cite{dwork2006calibrating} of the estimates in any range of values using the Count Sketch is at most $d$, since each HSDir contribution increases by at most 1 in at most $d$ values into the $d \cdot w$ Count Sketch table. Therefore, we can achieve $\epsilon$-differential privacy if we add, to each decrypted value, noise from a Laplace distribution with mean zero and variance $\xi \cdot d / \epsilon$, where $\xi$ is the number of decrypted intermediate results and $\epsilon$ the differential privacy parameter. However, doing so may result in the divide-and-conquer algorithm mis-estimating the range in which the median lies, and results in further mistakes in the final median estimate. (As discussed in Section~\ref{sec:diffPrivacy}, although we use $\epsilon$
to denote a parameter for both Count Sketch and differential privacy, it is clear from the context which one it relates to.)

\subsection{Implementation and Evaluation}

We implement and evaluate the proposed scheme aiming to: (i) estimate the trade-off between size of the sketch and the accuracy of the median computation, (ii) evaluate the cost of cryptographic computation and communication overheads, and (iii) assess the trade-off between the accuracy of the median and the quality of protection that may be achieved through the differentially private mechanism.

For our evaluation, we instantiate AH-ECC using the NIST-P224 curve as provided by the OpenSSL library and its optimizations by K{\"a}sper~\cite{kasper2012fast}. Our implementation of the cryptographic core of the private median scheme amounts to 300 lines of Python code using the \emph{petlib} OpenSSL wrapper\footnote{\url{https://github.com/gdanezis/petlib}}, and another 350 lines of Python include unit tests and measurement code. All experiments have been performed on a Xubuntu Trusty (Ubuntu 14.04.2 LTS) Linux VM, running on a 64 bit Windows 7 host (CPU i7-4700MQ, 2.4Ghz, 16GB RAM).
Our Python implementation is easily pluggable as part of the Tor infrastructure and does not require changes within the Tor (C-based) core functionalities.

We first illustrate the performance and accuracy of estimating the median using this technique with both sketch parameters $\epsilon$ and $\delta$ equal to either 0.25 or 0.05 against the London Atlas Dataset\footnote{\url{http://data.london.gov.uk/dataset/ward-profiles-and-atlas}} in Table~\ref{tab:stats} (see Appendix). The error rate is computed as the absolute value of difference between the estimated and true median divided by the true median.

Further results are presented on an experimental setup that uses as a reference problem the median estimation in a set of 1,200 sample values, drawn from a mixture distribution: 1,000 values from a Normal distribution with mean 300 and variance 25, and 200 values drawn from a Normal distribution with mean 500 and variance 200. This reference problem closely matches the settings of the Tor project both in terms of the range of vales (assumed to be within $[0, 1000]$) and the number of samples~\cite{elahi2014privex}.

\begin{figure}[t]
\centering
\includegraphics[width=0.9\columnwidth]{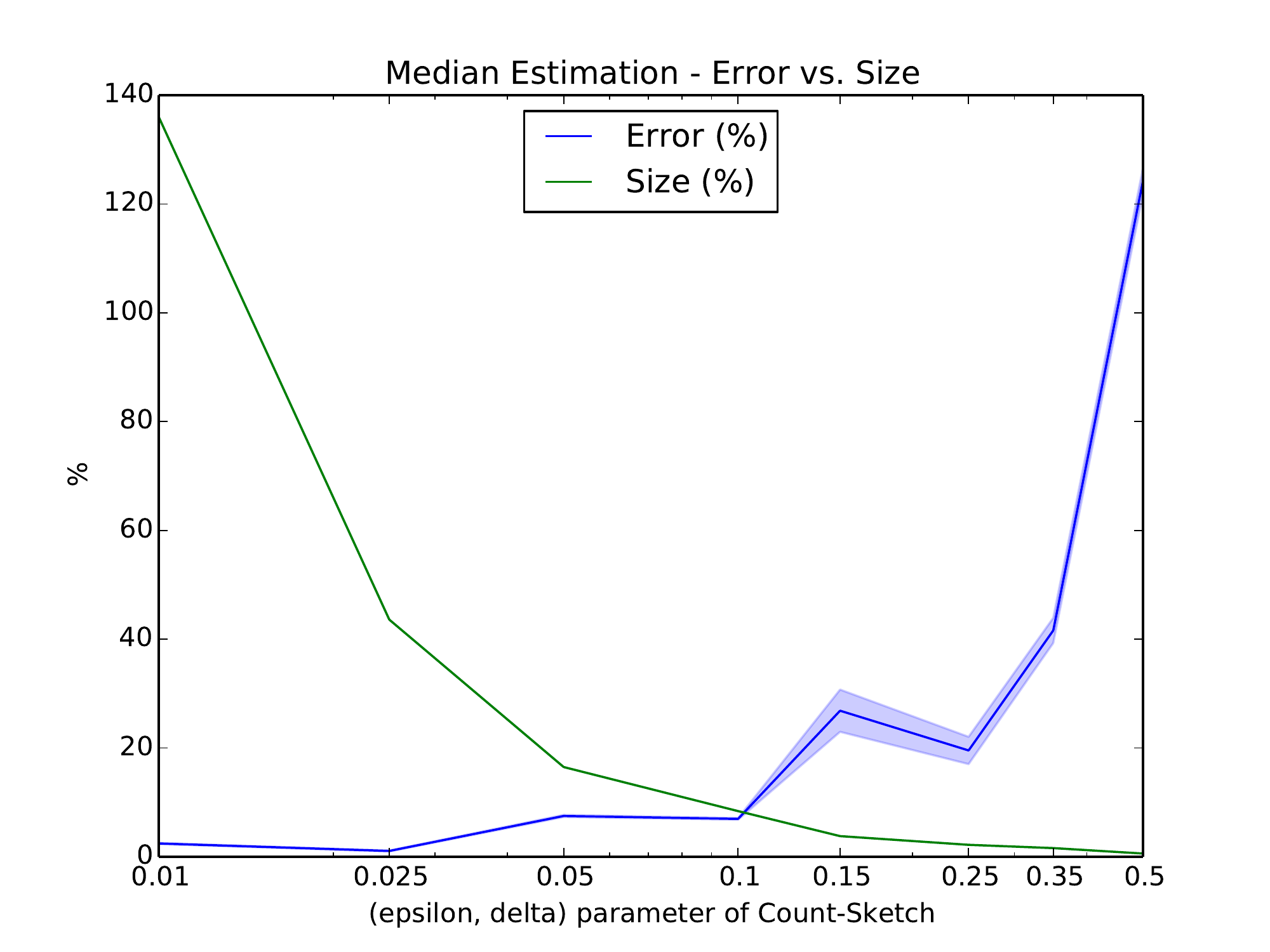}
\vspace{-0.2cm}
\caption{\label{fig:tor-sizes} Count Sketch size versus estimation quality.}
\vspace{-0.2cm}
\end{figure}

\descr{Quality vs. Size.} Figure~\ref{fig:tor-sizes} illustrates the trade-off between the quality of the estimation of the median algorithm and the size overhead of the Count Sketch. The size overhead (green slim line) is computed as the number of encrypted elements in the sketch as compared with the number of elements in the range of the median (1,000 for our reference problem). The estimation accuracy (blue broader line) is represented as the fraction of the absolute deviation of the estimate from the real value over the real sample median (light blue region represents the standard deviation of the mean over 40 experiments for each datapoint). Thus both qualities can be represented as percentages. 

The trade off between the size of the sketch and the accuracy of the estimate is evident: as the sketch size reaches a smaller fraction of the total possible number of values, the error becomes larger than the range of the median. Thus, Count Sketch with parameters $ \epsilon, \delta < 0.025  $ are unnecessary, since they do not lead to a reduction of the information that needs to be transmitted from each client to the authorities; conversely, for $ 0.15 < \epsilon, \delta $ the estimate of the median deviates by more than 20\% of its true value making it highly unreliable.

For all subsequent experiments, we consider a Count Sketch with values $\epsilon = \delta = 0.05$, leading to $d = 3$ and $w = 55$. As outlined in Figure~\ref{fig:tor-sizes}, this represents a good trade-off between the size of the Count Sketch ($16.5\%$ of transmitting all values) and the error. 

\descr{True Size and Performance.} When implemented using NIST-P224 curves, the reference Count Sketch may be serialized in 10,898 bytes. Each Count Sketch takes 0.001 sec to encrypt at each HSDir, and it takes 1.456 seconds to aggregate 1,200 sketches at each authority (0.001 sec per sketch). As expected, from the range of the reference problem, 10 decryption iterations are sufficient to converge to the median (therefore $\xi = 10$). The number of homomorphic additions for each decryption round is linear in the range of the median %
and their total computational cost is the same order of magnitude as a full Count Sketch encryption. It is clear from these figures that the computational overhead of the proposed technique is eminently practical, and the bandwidth overhead acceptable.

\begin{figure}[t]
\centering
\includegraphics[width=0.9\columnwidth]{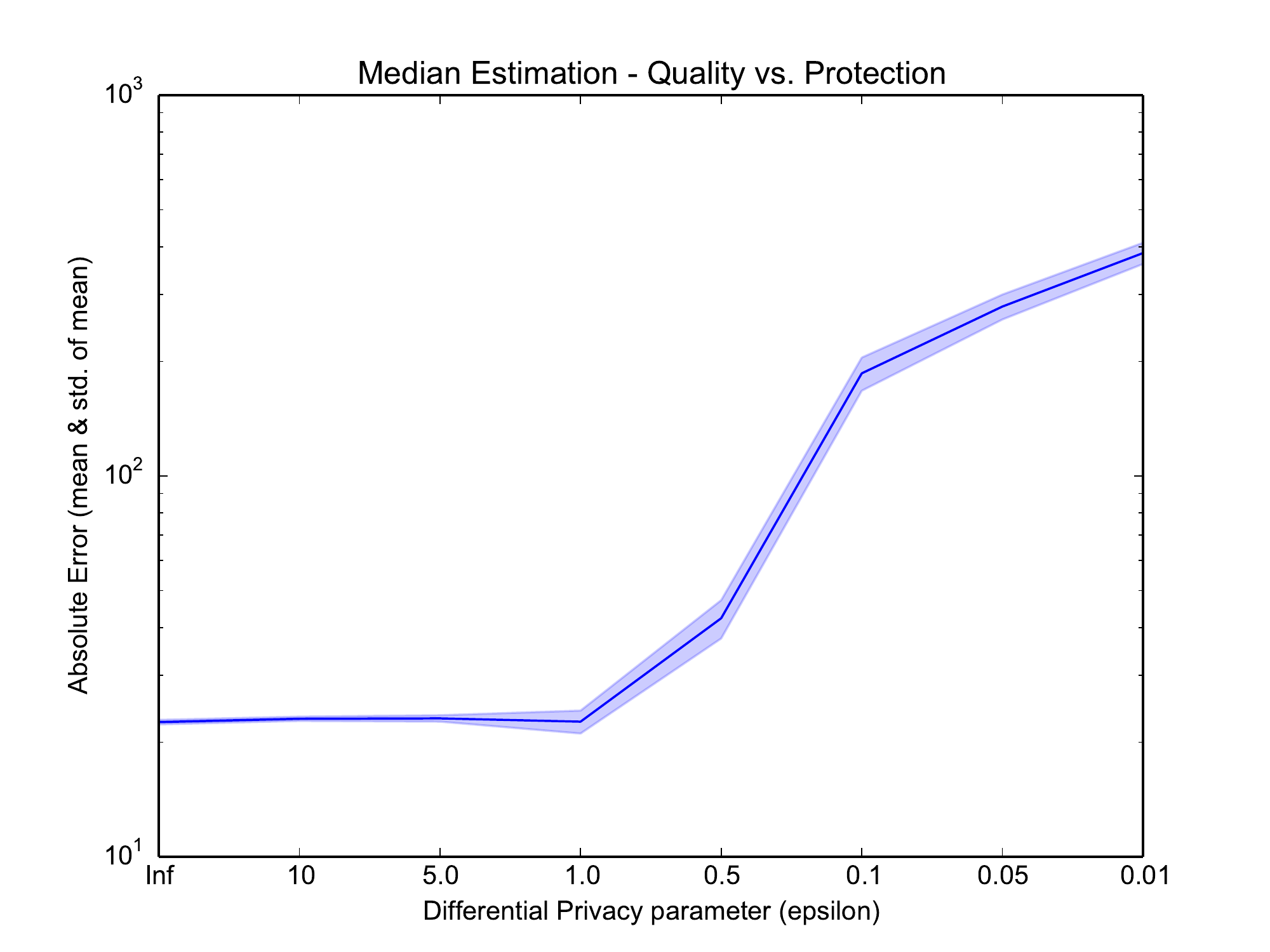}
\vspace{-0.2cm}
\caption{\label{fig:tor-quality} Quality versus differential privacy protection.}
\vspace{-0.2cm}
\end{figure}

\descr{Quality vs. Differential Privacy Protection.} Figure~\ref{fig:tor-quality} illustrates the trade-off between the quality of the median estimation and the quality of differential privacy protection. The x-axis represents the $\epsilon$ parameter of the differentially private system, and the y-axis the absolute error between the estimate and the true sample median. Differential privacy with parameter $\epsilon = 0.5$ can be provided without significantly affecting the quality of the median estimate. However, for $\epsilon < 0.5$ the volume of the error grows exponentially (note the log scale of the x-axis). While the exact value of a 
meaningful $\epsilon$ parameter is often debated in the literature, we conclude that the mechanism only provides a limited degree of protection, and no ability to readily tune up protection: utility degrades very rapidly as the security parameter $\epsilon$ decreases.

\section{Related Work}
\label{sec:work}
This section reviews prior work on privacy-preserving techniques applied to data aggregation, recommender systems,  machine learning, participatory sensing, as well as efficient data structures for succinct representation. %

\subsection{Privacy-Preserving Aggregation}
\label{sec: data-aggr} 
Kursawe et al.~\cite{Kursawe:2011} introduce a few cryptographic constructions to aggregate energy consumptions in the context of smart metering, relying on Diffie-Hellman, %
bilinear maps, and a {\em ``low overhead''} protocol where meters' encryption keys sum up to zero. 
Our schemes for the private recommender system (Section~\ref{sec:service}) 
and location prediction (Section~\ref{sec:smartphone}) rely on 
a protocol inspired by \cite{Kursawe:2011}'s ``low overhead'' protocol, but perform private aggregation using succinct data representation rather than the raw inputs. Using Count-Min Sketch~\cite{cormode2005improved}, we reduce computation and communication overhead incurred by each user from linear to logarithmic in the size of the input. We also show how to recover from node failures, i.e., in our schemes, the aggregator can still retrieve the statistics (and train models) even when a subset of users go offline or fail to report data.

Castelluccia et al.~\cite{Mobiquitous05} propose a new homomorphic encryption to allow intermediate wireless sensor nodes to aggregate encrypted data gathered from other nodes.  Shi et al.~\cite{shi2011privacy} combine private aggregation with differential privacy supporting the aggregation of encrypted perturbed readings reported by the meters. Individual amounts of random noise cancel each other out during aggregation, except for a specific amount that guarantees computational differential privacy. Their protocol is also so that encryption keys sum up to zero but, unlike ours, requires solving a discrete logarithm and the presence of a trusted dealer. 
Jawurek et al.~\cite{jawurek2012fault} propose a privacy-friendly aggregation scheme with robustness against missing user inputs, by including additional authorities that facilitate the protocol but do not learn any secrets or inputs.
However, at least one of the authorities has to be honest, i.e., if all collude, the protocol does not provide any privacy guarantee.
Chan et al.~\cite{chan2012privacy} also provides fault tolerance by extending 
\cite{shi2011privacy}'s protocol, however, with a poly-logarithmic penalty. 
Additional, more loosely related, private aggregation schemes include~\cite{Mobiquitous05,erkin2012private,bilogrevic2014s}. 

A combination of homomorphic encryption and differential privacy has  been explored by Chen et al.~\cite{chen2012towards}, allowing third parties to gather web analytics.
Users encrypt their data using the data aggregator public key and send them to a proxy, who adds noise to the ciphertexts and forwards the results to the data aggregator. The latter computes the aggregates after decrypting each individual contribution.
However, this scheme introduces a large overhead both in terms of communication (one KB per single bit of user data) and computation (one public key operation per single bit).
In the same line of work, Akkus et al.~\cite{akkus2012non} propose a system providing differential privacy guarantees. 
Their scheme scales better than~\cite{chen2012towards} as it requires users to encrypt fewer bits per query, but still relies on expensive public-key crypto operations.
In~\cite{chen2013splitx}, the authors propose a scheme based on a similar trust model as~\cite{chen2012towards} but with an enhanced scalability by using simple exclusive-or (XOR) operations rather than public key operations. However, their proposal still relies on honest-but-curious servers that do not collude with each other.

Erlingsson et al.~\cite{erlingsson2014rappor} introduce RAPPOR, which enables the collection of browser statistics on values and strings provided by a large number of clients (e.g. homepage settings, running processes, etc.), including categories, frequencies, and histograms.
RAPPOR supports privacy-preserving data-collection mechanism by relying on randomized responses via input perturbation, aiming to guarantee local differential privacy for  individual reports.
This, however, requires millions of users in order to obtain approximate answers to queries.

Finally, Elahi et al.~\cite{elahi2014privex} present a protocol for privately computing mean statistics on Tor traffic.
They introduce two ad-hoc protocols relying, respectively, on secret sharing and distributed decryption. By contrast, our application for gathering private statistics for Tor enables the computation of the median statistics on traffic generated by Tor hidden services -- which constituted an open problem~\cite{goulethidden} -- by relying on additively homomorphic encryption and differential privacy. %

\subsection{Privacy-preserving Recommender Systems}
McSherry and Mironov~\cite{mcsherry2009differentially} propose a privacy-preserving recommender system that relies on trusted computing, while Ciss{\'e}e and Albayrak~\cite{cissee2007agent} use differential privacy to add privacy guarantees to a few algorithms presented during the Netflix Prize competition.
Our private recommender system differs from theirs as we do not rely on trusted computing or differential privacy, but leverage a privacy-friendly aggregation cryptographic protocol and Count-Min Sketch.

Homomorphic encryption based techniques have also been used to perform other machine learning operations on encrypted data, including matrix factorization \cite{nikolaenko2013privacy}, linear classifiers \cite{bos2014private,graepel2013ml}, and decision trees \cite{bost2014machine}. 
Building a cloud-based model from multiple user datasets has been also addressed in \cite{lopez2012fly}, which explores the feasibility of Fully Homomorphic Encryption (FHE) based techniques.
However, at the moment, FHE operations are still prohibitively expensive. %

\subsection{Participatory Sensing}
Mood et al.~\cite{mood2014reuse} propose a privacy-preserving participatory sensing application which allows users to locate nearby friends without disclosing exact locations, via secure function evaluation~\cite{yao1982protocols}, but do not address the problem of scaling to large streams/number of users. De Cristofaro and Soriente~\cite{de2013extended} introduce a privacy-enhanced distributed querying infrastructure for participatory and urban sensing systems.
Work in~\cite{mobisys08} and~\cite{comcom} provide either $k$-anonymity~\cite{k-anonymity} and $l$-diversity \cite{l-diversity} to guarantee anonymity of users through Mix Network techniques~\cite{MixNet}. However, their techniques are not provably-secure and they only provide partial confidentiality.
Then,~\cite{ganti2008} suggest data perturbation in a known community for computing statistics and protecting anonymity.
Trusted Platform Modules (TPMs) are instead used in~\cite{gilbert2010} and~\cite{dua2009} to protect integrity and authenticity of user contents.

In a way, we also address the problem of participatory sensing privacy by proposing a scalable and provable secure technique for collecting user-generated streams of data involving a large number of users.

\subsection{Privacy and Succinct Data Representation}
Mir et al.~\cite{mir2011pan} present an efficient scheme guaranteeing differential privacy of data analyses (even when the internal memory of the algorithm may be compromised), using a data structure similar to the Count-Min Sketch to estimate heavy hitters. 
Work in~\cite{hsu2012distributed, chan2012differentially} address the problem of finding heavy hitters' histograms while preserving privacy using 
a differentially private protocol. 
Then,~\cite{bassilylocal} addresses the case where individual users randomize their own data and then send differentially private reports to an untrusted server handling reports aggregation.  
Other proposals combine differential privacy and Count-Min Sketch to obtain aggregate information about vehicle traffic~\cite{Monreale2013} as well as summaries of sparse databases~\cite{cormode2012differentially}.

Ashok et al.~\cite{ashok2014scalable} present a privacy-preserving protocol for computing the set-union cardinality among several parties using Bloom filters~\cite{bloom1970space}. However, their proposal is insecure, as shown by~\cite{egertprivately}, who also introduces a novel Bloom filter based protocol for set-union and set-intersection cardinality. 
Lin et al.~\cite{lin2012efficient} improve the performance of \cite{narayanan2011location}'s protocol for private proximity testing by reducing the problem to simple equality testing (instead of the more expensive private-preserving threshold set intersection). They use a concise representation of ``location tags'', by generating, via shingling, concise sketches---in their context, short strings representing the set of broadcast messages received. 

In summary, to the best of our knowledge, our work is the first to show how to combine Count-Min Sketch and privacy-friendly data aggregation to build a private estimated model used for recommendations as well as prediction of future locations. Also, our scheme for Tor hidden services statistics, which combines Count Sketch, additively homomorphic threshold decryption, and differential privacy, is the first to tackle the problem of efficiently computing the median statistics.

\section{Conclusion}\label{sec:conclusion}
This paper presented efficient techniques for privately and efficiently collecting statistics by relying on private data aggregation protocols and succinct data structures.
These allowed us to reduce the communication and computation complexity incurred by each data source from linear to logarithmic in the size of the input but only introduced a limited, upper-bounded error in the quality of the statistics.

Our techniques support different trust, robustness, and deployment models and can be applied to a number of interesting real-world problems where aggregate statistics can be used to train models. We presented the design and deployment of a private recommender system for streaming services and a private location prediction service. Our server-side implementation as a JavaScript web application allows developers to easily incorporate it in their projects, while user-side is supported both in the browser (thus requiring users to install no additional software) and in Android.
We also designed and implemented (in Python) a scheme for computing the median statistics of Tor hidden services in a privacy-friendly way.

As part of future work, we plan to apply our private recommender system to the BBC news apps for Android, conduct a test deployment of the private location prediction service with a local mass transit operator, and extend our protocols to privately consolidate data shared by different sources~\cite{freudiger2015controlled}. We are also working on releasing a comprehensive framework supporting large-scale privacy-preserving aggregation {\em as a service}. 

\descr{Acknowledgements.} 
We would like to thank Chris Newell and Michael Smethurst from the BBC and Aaron Johnson from US Naval Research Labs for motivating our work, respectively, on privacy-preserving recommendation and median statistics in Tor.
We are also grateful to Mirco Musulesi, Licia Capra, and Apostolos Pyrgelis for providing feedback and useful comments. Luca Melis and Emiliano De Cristofaro are supported by a Xerox's University Affairs Committee award on ``Secure Collaborative Analytics'' and ``H2020-MSCA-ITN-2015'' Project Privacy{\&}Us (ref.\ 675730). George Danezis is supported in part by EPSRC Grant ``EP/M013286/1'' and H2020 Grant PANORAMIX (ref.\ 653497).
%

%
%\bibliographystyle{abbrv}
%\bibliography{bibfile}

%
%

\begin{table*}[t]
\centering
\footnotesize
\begin{tabular}{lrrrrr}
\hline
{} &  Median ($\epsilon, \delta = 0.25$) &  Error (\%) &  Median ($\epsilon, \delta = 0.05$) &  Error (\%) &   Truth \\
\hline
Population - 2015                                  &        15143.2 &       11.3 &        13215.4 &        2.8 & 13600.0 \\
Children aged 0-15 - 2015                          &         2970.8 &       12.1 &         2627.6 &        0.8 &  2650.0 \\
Working-age (16-64) - 2015                         &         9592.0 &        2.0 &         8843.2 &        5.9 &  9400.0 \\
Older people aged 65+ - 2015                       &         1284.6 &       11.4 &         1345.0 &        7.2 &  1450.0 \\
\% All Children aged 0-15 - 2015                    &           21.9 &       10.7 &           20.1 &        1.3 &    19.8 \\
\% All Working-age (16-64) - 2015                   &           70.7 &        5.0 &           68.8 &        2.2 &    67.3 \\
\% All Older people aged 65+ - 2015                 &           15.2 &       37.1 &           12.0 &        7.8 &    11.1 \\
Mean Age - 2013                                    &           38.6 &        8.8 &           36.9 &        3.8 &    35.5 \\
Median Age - 2013                                  &           37.7 &       10.8 &           35.7 &        5.1 &    34.0 \\
Area - Square Kilometres                           &            0.6 &       68.1 &            1.6 &       16.9 &     1.9 \\
Population density (persons per sq km) - 2013      &        10231.3 &       44.8 &         5792.9 &       18.0 &  7067.0 \\
\% BAME - 2011                                      &           45.6 &       26.3 &           35.7 &        1.0 &    36.1 \\
\% Not Born in UK - 2011                            &           40.1 &        7.6 &           40.1 &        7.6 &    37.3 \\
\% English is First Language of no one in househ... &           16.9 &       41.7 &           11.8 &        0.9 &    11.9 \\
General Fertility Rate - 2013                      &           73.3 &       14.4 &           66.8 &        4.1 &    64.1 \\
Male life expectancy -2009-13                      &           84.1 &        5.7 &           79.6 &        0.0 &    79.6 \\
Female life expectancy -2009-13                    &           87.0 &        3.5 &           84.9 &        0.9 &    84.1 \\
Rate of All Ambulance Incidents per 1,000 popul... &           52.5 &       54.9 &           98.6 &       15.3 &   116.3 \\
Rates of ambulance call outs for alcohol relate... &            0.1 &       78.0 &            1.0 &       74.0 &     0.6 \\
Number Killed or Seriously Injured on the roads... &            3.0 &        1.3 &            3.5 &       16.7 &     3.0 \\
In employment (16-64) - 2011                       &         6532.8 &        7.0 &         5843.7 &        4.2 &  6103.0 \\
Employment rate (16-64) - 2011                     &           68.5 &        2.0 &           70.8 &        1.3 &    69.9 \\
Rate of new registrations of migrant workers - ... &           42.9 &       10.7 &           34.5 &       11.1 &    38.8 \\
Number of properties sold - 2013                   &          169.3 &        1.4 &          149.8 &       10.3 &   167.0 \\
Modelled Household median income estimates 2011/12 &        31802.6 &        2.2 &        29589.3 &        9.0 & 32509.0 \\
Number of Household spaces - 2011                  &         5619.1 &        5.4 &         5025.9 &        5.7 &  5332.0 \\
\% detached houses - 2011                           &            2.4 &       44.7 &            1.6 &       62.2 &     4.3 \\
\% semi-detached houses - 2011                      &           29.0 &       70.6 &           16.7 &        1.5 &    17.0 \\
\% terraced houses - 2011                           &           29.4 &       39.8 &           21.1 &        0.6 &    21.0 \\
\% Flat, maisonette or apartment - 2011             &           53.1 &       15.1 &           49.7 &        7.9 &    46.1 \\
\% Households Owned - 2011                          &           57.3 &       18.4 &           53.3 &       10.2 &    48.4 \\
\% Households Social Rented - 2011                  &           26.0 &       27.5 &           19.9 &        2.4 &    20.4 \\
\% Households Private Rented - 2011                 &           30.9 &       26.5 &           26.6 &        9.1 &    24.4 \\
\% dwellings in council tax bands A or B - 2011     &           21.2 &       79.9 &           10.4 &       12.2 &    11.8 \\
\% dwellings in council tax bands C, D or E - 2011  &           63.7 &        7.5 &           71.6 &        3.9 &    68.9 \\
\% dwellings in council tax bands F, G or H - 2011  &            0.3 &       96.7 &            1.4 &       82.6 &     8.1 \\
Claimant Rate of Incapacity Benefit - 2014         &            1.8 &       80.0 &            0.9 &       10.0 &     1.0 \\
Claimant Rate of Income Support - 2014             &            4.4 &      119.6 &            2.3 &       16.8 &     2.0 \\
Claimant Rate of Employment Support Allowance -... &            6.9 &       65.3 &            4.7 &       13.0 &     4.2 \\
Rate of JobSeekers Allowance (JSA) Claimants - ... &            5.0 &       34.6 &            3.1 &       16.6 &     3.7 \\
\% dependent children (0-18) in out-of-work hous... &           22.2 &       19.6 &           19.1 &        2.6 &    18.6 \\
\% of households with no adults in employment wi... &            8.7 &       67.2 &            5.3 &        1.2 &     5.2 \\
\% of lone parents not in employment - 2011         &           51.9 &       11.2 &           47.5 &        1.6 &    46.7 \\
(ID2010) - Rank of average score (within London... &          366.3 &       17.4 &          301.6 &        3.3 &   312.0 \\
(ID2010) \% of LSOAs in worst 50\% nationally - 2010 &           -6.4 &      107.7 &           99.2 &       19.5 &    83.0 \\
Average GCSE capped point scores - 2013            &          369.0 &        6.0 &          349.4 &        0.4 &   348.0 \\
Unauthorised Absence in All Schools (\%) - 2013     &            1.7 &       53.5 &            0.8 &       26.2 &     1.1 \\
\% with no qualifications - 2011                    &           20.8 &       19.1 &           18.8 &        7.2 &    17.5 \\
\% with Level 4 qualifications and above - 2011     &           44.4 &       25.1 &           39.1 &       10.1 &    35.5 \\
A-Level Average Point Score Per Student - 2012/13  &          715.3 &        5.7 &          668.4 &        1.3 &   676.9 \\
A-Level Average Point Score Per Entry; 2012/13     &          215.0 &        3.1 &          210.8 &        1.1 &   208.5 \\
Crime rate - 2013/14                               &         1163.6 &     1598.7 &           47.8 &       30.3 &    68.5 \\
Violence against the person rate - 2013/14         &            1.2 &       92.5 &           10.5 &       35.6 &    16.3 \\
Robbery rate - 2013/14                             &            1.6 &       31.8 &            0.1 &       94.7 &     2.3 \\
Theft and Handling rate - 2013/14                  &           -3.5 &      113.7 &           11.4 &       55.6 &    25.6 \\
Criminal Damage rate - 2013/14                     &            9.1 &       43.8 &            5.9 &        6.6 &     6.3 \\
Drugs rate - 2013/14                               &           -9.3 &      321.4 &            2.8 &       33.8 &     4.2 \\
\% area that is open space - 2014                   &           30.1 &       28.3 &           19.3 &       17.9 &    23.5 \\
Cars per household - 2011                          &            1.6 &       99.4 &            0.5 &       35.0 &     0.8 \\
Average Public Transport Accessibility score - ... &            6.8 &       99.6 &            4.4 &       28.9 &     3.4 \\
\% travel by bicycle to work - 2011                 &           12.0 &      343.9 &            3.0 &       12.5 &     2.7 \\
Turnout at Mayoral election - 2012                 &           38.1 &       11.5 &           35.0 &        2.3 &    34.2 \\
\hline
\end{tabular}
\caption{Median estimation with $22$ ciphertexts ($d=2$, $w=11$, $\epsilon, \delta = 0.25$) and $165$ ciphertexts ($d=3$, $w=55$, $\epsilon, \delta = 0.05$) on the London Atlas Dataset.}\label{tab:stats}
\end{table*}

\end{document}